\documentclass{aa}  
\usepackage{graphicx,natbib,url,twoopt}
\usepackage{graphicx,url,twoopt}
\usepackage[varg]{txfonts}           
\usepackage{hyperref}                

\usepackage[caption = false]{subfig}
\usepackage{graphicx}
\makeatletter
\newcommand{\bibnote}[2]{\@namedef{#1note}{#2}}
\newcommand{\biblink}[2]{\@namedef{#1link}{#2}}
\makeatother

\begin{document}
\title{CO and HCN isotopologue ratios in the outflows of AGB stars}
\author{M. Saberi\inst{\ref{1},\ref{2},\ref{3}}
        \and H. Olofsson\inst{\ref{1}}
        \and W. H. T. Vlemmings\inst{\ref{1}}
       \and  E. De Beck\inst{\ref{1}}
        \and  T. Khouri\inst{\ref{1}}
      \and  S. Ramstedt\inst{\ref{4}}
      }
\institute{ Department of Space, Earth and Environment, Chalmers University of Technology, Onsala Space Observatory, 439 92 Onsala, Sweden \label{1}
\\ \email{maryam.saberi@chalmers.se}
\and
Rosseland Centre for Solar Physics, University of Oslo, P.O. Box 1029 Blindern, NO-0315 Oslo, Norway
\label{2}
\and
Institute of Theoretical Astrophysics, University of Oslo, P.O. Box 1029 Blindern, NO-0315 Oslo, Norway
\label{3}
\and
Department of Physics and Astronomy, Division of Astronomy \& Space Physics, Uppsala University, POBox 516, 751 20 Uppsala, Sweden
\label{4}
 }
\date{}

\abstract 
{Isotopologue line intensity ratios of circumstellar molecules have been widely used to trace the photospheric elemental isotopic ratios of evolved stars. However, depending on the molecular species and the physical conditions of the environment, the isotopologue ratio in the circumstellar envelope (CSE) may deviate considerably from the stellar atmospheric value.}
{In this paper, we aim to examine how the $^{12}$CO/$^{13}$CO and H$^{12}$CN/H$^{13}$CN abundance ratios vary radially due to chemical reactions in the outflows of asymptotic giant branch (AGB) stars and the effect of excitation and optical depth on the resulting line intensity ratios. We study both carbon-rich (C-type) and oxygen-rich (O-type) CSEs.}
{We performed chemical modeling to derive radial abundance distributions of our selected species in the CSEs over a wide range of mass-loss rates ($10^{-8}<\dot{M}<10^{-4} M_{\odot}$ yr$^{-1}$). We used these as input in a non-local thermodynamic equilibrium (non-LTE) radiative transfer code to derive the line intensities of several ground-state rotational transitions. 
We also test the influence of stellar parameters, physical conditions in the outflows, the intensity of the interstellar radiation field, and the importance of considering the chemical networks in our model results.}
{We quantified deviations from the atmospheric value for typical outflows. We find that the circumstellar value of $^{12}$CO/$^{13}$CO can deviate from its atmospheric value by up to 25-94\,$\%$ and 6-60\,$\%$ for C- and O-type CSEs, respectively, in radial ranges that depend on the mass-loss rate.
We show that variations of the intensity of the interstellar radiation field and the gas kinetic temperature can significantly influence the CO isotopologue abundance ratio in the outer CSEs of both C-type and O-type.
On the contrary, the H$^{12}$CN/H$^{13}$CN abundance ratio is stable throughout the CSEs for all tested mass-loss rates.
The radiative transfer modeling shows that the integrated line intensity ratio $I_{\rm ^{12}CO}/I_{\rm ^{13}CO}$ of different rotational transitions varies significantly for stars with mass-loss rates in the range from $10^{-7}$ to $10^{-6}$\,$M_{\odot}$\,yr$^{-1}$ due to combined chemical and excitation effects. In contrast, the excitation conditions for the HCN isotopologues are the same for both isotopologues.}
{We demonstrate the importance of using the isotopologue abundance profiles from detailed chemical models as inputs to radiative transfer models in the interpretation of isotopologue observations. Previous studies of circumstellar CO isotopologue ratios are based on multi-transition data for individual sources and it is difficult to estimate the errors in the reported values due to assumptions that are not entirely correct according to this study. If anything, previous studies may have overestimated the circumstellar $^{12}$CO/$^{13}$CO abundance ratio. The use of the HCN molecule as a tracer of C isotope ratios is affected by fewer complicating problems, but we note that the corrections for high optical depths are very large in the case of high-mass-loss-rate C-type CSEs; and in O-type CSEs the H$^{13}$CN lines may be too weak to detect.}

\keywords{Stars: abundances -- Stars: AGB -- Stars: circumstellar matter -- Ultraviolet: stars}
\maketitle

\section{Introduction}\label{Introduction}

In the late stages of their evolution, stars with initial masses of 0.8--8 $M_\odot$ evolve along the asymptotic giant branch (AGB), where they lose up to 80\,\% of their mass due to strong stellar winds \citep{Hofner18}. 
As a consequence, a large circumstellar envelope (CSE) of dust and gas forms around the AGB star.
Observations of the molecular compositions of these CSEs provide information on the CSE chemistry, on the recycled material, and, consequently, on their stellar properties.
For instance, measuring the atomic isotopic ratio $^{12}$C/$^{13}$C provides information on the nucleosynthesis of the star, its phase of evolution, and its properties.
This is because the abundances of different C isotopes vary due to stellar nucleosynthesis and mixing processes, and the effect of these processes, in turn, depends on the stellar mass and the evolutionary phase \citep[e.g.,][]{Karakas14}.

Estimating the atomic isotope abundances in AGB stars from observational data is challenging. One alternative possibility is to use circumstellar isotopologue ratios. Considering the high temperature in the stellar atmosphere, we expect the initial isotopologue abundance ratios of molecules containing the desired atomic isotopes to directly trace the atomic isotopic ratios.
However, in order to use the former as tracers of the latter, it is crucial to consider all possible processes that can alter the molecular isotopologue ratios in the CSE.

The circumstellar $^{12}$CO/$^{13}$CO ratio has been extensively used to estimate the stellar $^{12}$C/$^{13}$C ratio \citep[e.g.,][]{Schoier00, Milam09, Ramstedt14}.
However, there are at least two known processes that can lead to a varying CO isotopologue ratio through a CSE: isotopologue-selective photodissociation through lines by UV radiation and chemical fractionation.
The UV photodissociation rate of CO depends on many factors such as the intensity of the local radiation field, as well as CSE properties such as H$_2$ density, CO isotopologue column density, gas kinetic temperature, and the clumpiness of the CSE \citep[e.g.,][]{Saberi19}. Generally, since $^{12}$CO is more abundant than $^{13}$CO, its shielding efficiency is higher than that of the less abundant isotopologue. This leads to an increase in the $^{12}$CO/$^{13}$CO ratio. The CO fractionation reaction is given by \cite{Watson76},
\begin{equation}
\rm ^{13}C^+ + ^{12}CO \:\: \rightleftharpoons \:\: ^{12}C^+ + ^{13}CO + \Delta E,
\end{equation}
where $\Delta$E corresponds to a temperature of 35\,K. At high temperatures, the forward and backward reaction rates are equal. However, at low temperatures ($T_{\rm kin}$\,$<$\,100\,K), the backward reaction rate decreases, resulting in a one-way channel which favours creation of $^{13}$CO from $^{12}$CO. Therefore, at low temperatures, this can lead to a lower $^{12}$CO/$^{13}$CO ratio. 
It has been argued that these two processes, isotopologue-selective photodissociation and fractionation reaction, neutralize each other in the outer CSE, leading to a constant $^{12}$CO/$^{13}$CO ratio equal to the initial value \citep[e.g.,][]{Mamon88,Ramstedt14}.
However, there is considerable uncertainty concerning the relative efficiencies of these processes.

Also HCN, another abundant molecule in AGB CSEs, can be used to trace the atomic C isotope ratio. 
In this case, there are no known processes that can significantly selectively alter the isotopologue abundances since this molecule is photodissociated in the continuum and chemical fractionation is not efficient. Hence, the H$^{12}$CN/H$^{13}$CN ratio might be a more reliable tracer of the $^{12}$C/$^{13}$C ratio, provided one accounts correctly for the optical thickness of the emission, which can be substantial in carbon-rich CSEs formed by high mass-loss rates.
In the case of R Scl, \cite{Saberi17} showed that the H$^{12}$CN/H$^{13}$CN ratio is in fact in agreement with the photospheric $^{12}$C/$^{13}$C ratio, while the $^{12}$CO/$^{13}$CO ratio varies by a factor of three through the CSE \citep{Vlemmings13}.

In this paper, we aim to investigate variations of the $^{12}$CO/$^{13}$CO and H$^{12}$CN/H$^{13}$CN ratios in AGB CSEs of both C-type (associated with carbon stars with C/O\,>\,1) and O-type (associated with M-type stars with C/O\,<\,1) over a wide range of mass-loss rates. We perform a detailed chemical analysis and derive the relevant abundance distributions for several different initial conditions. The derived abundance distributions are used as input in radiative transfer models to derive the line intensities of several rotational transitions of both CO and HCN. 
This provides a good estimate of the reliability of the CO and HCN isotopologue ratios as measures of the carbon isotopic ratio in the CSEs of AGB stars.

\section{Circumstellar chemistry}\label{Code}                              

In order to model the circumstellar chemistry we used an extended version of the publicly available chemical  code rate13-cse\footnote{http://udfa.ajmarkwick.net/index.php?mode=downloads} for CSEs \citep{McElroy13}. 
The code assumes a spherically symmetric CSE formed by a constant mass-loss rate $\dot{M}$. It expands with a constant velocity $V_{\rm exp}$. The H$_2$ density falls off as $1/r^2$ where $r$ measures the distance from the central star.
We upgraded the chemical network of the code to include isotopic chemistry, see Sect.\ref{Chemical Network}.
We also upgraded the calculations of the CO photodissociation rate as detailed in \cite{Saberi19}.

\subsection{Chemical network}\label{Chemical Network}

We incorporate an extended version of the chemical network from the UMIST Database for Astrochemistry which includes the $^{13}$C and $^{18}$O isotopes, all corresponding molecular isotopologues, their chemical reactions, and the properly scaled reaction rate coefficients from \cite{Rollig13}.
The chemical network includes 933 species containing 351 neutral species, 572 cations and 10 anions. We consider 15108 gas-phase reactions. The type of reactions, number of reactions, and reaction rates are listed in Table \ref{reactions}. Detailed explanations of the reaction types and calculations of the reaction rates can be found in \cite{McElroy13} and \cite{Rollig13}.

The initial abundances of parent species are listed in Table~\ref{ParentS}, and the reasons for their choices are given in Appendix~\ref{s:chemistry}.
We assume the fractional abundance of $^{12}$CO/H$_2$ to be $8\times10^{-4}$ and $5\times10^{-4}$ for C-type and O-type CSEs, respectively. This corresponds to almost complete formation of CO assuming solar abundances of C and O; see Appendix~\ref{s:chemistry}.
The H$^{12}$CN/H$_2$ values are taken from the observationally determined median values of \cite{Schoier13}, $3\times10^{-5}$ and $1\times10^{-7}$ for samples of C-type and O-type CSEs, respectively.
The initial abundance of $^{13}$C-bearing molecules are scaled down from their $^{12}$C-bearing isotopologue by factors of 34 and 13 for C-type and O-type CSEs, respectively. These are the median $^{12}$CO/$^{13}$CO values for a sample of 19 C-type and 19 M-type AGB stars reported by \cite{Ramstedt14}.

\begin{table*}[t]
  \centering
  \caption{Type, number, and reaction rates of all reactions in the chemical network of the CSE model.}
  \begin{tabular}{lllcl}
\hline\hline
Reaction type & Num & Reaction rate \\
\hline
Associative Detachment (AD) & 136 &$k_1 = \alpha (\frac{T}{300})^\beta \exp(\frac{-\gamma}{T})  \:\:  (\rm cm^3 s^{-1})$\\
Collisional Dissociation (CD) & 44 &$ k_1$\\
Charge Exchange (CE) & 1536 & $k_1$\\
Dissociative Recombination (DR) & 1273 & $k_1$\\
Ion-Neutral (IN) & 8737 & $k_1$\\
Mutual Neutralisation (MN) & 39 & $k_1$\\
Neutral-Neutral (NN) & 1986 & $k_1$\\
Radiative Association (RA) & 294 & $k_1$\\
Radiative Electron Attachment (REA) & 8 & $k_1$\\
Radiative Recombination (RR) & 28 & $k_1$\\
Cosmic-Ray Proton (CP) & 16 & $k_2 = \alpha\:\: (s^{-1})$ \\       
Cosmic-Ray Photon (CR) & 443 & $k_3 =  \alpha (\frac{T}{300})^\beta \frac{\gamma}{1-\omega} \:\:\:\: (\rm s^{-1})$ \\
Photo Process (PH) & 568 & $k_4 = \alpha \exp(-\gamma A_{\rm V}) \:\:\:\: (\rm s^{-1}) $\\  
\hline
  \end{tabular}
  \label{reactions}
 \tablefoot{Reaction rates applicable to different reaction types follow the analytical prescriptions $k_1,k_2,k_3,k_4$. In all of these, $T$ is the gas temperature; the other parameters are defined as follows: $k_1$:  $\alpha$ is the pre-exponential factor, $\beta$ indicates the temperature dependence of the rate coefficient and $\gamma$ is the activation energy of the reaction; $k_2$: $\alpha$ is the cosmic-ray ionisation rate; $k_3$: $\alpha$ is the cosmic-ray ionisation rate, $\gamma$ is the probability per cosmic-ray ionisation that the photo-reaction takes place and $\omega$ is the dust-grain albedo; $k_4$: $\alpha$ is the photodissociation rate in the unshielded region, $A_V$ is the extinction by the interstellar dust at visible wavelength and $\gamma$ is a parameter to take into account the increased extinction at UV wavelengths compared to the visible.}
\end{table*}


\subsection{CSE properties}\label{CSE properties}

We considered five reference models for each chemical type of CSE with mass-loss rates of $10^{-8}, 10^{-7},..,10^{-4}$ $M_{\odot}$ yr$^{-1}$, and a constant expansion velocity $V_{\rm exp}$\,=\,15\,km\,s$^{-1}$.
Although it is known that the wind is accelerating, we assumed a constant velocity since an implementation of an accelerating wind in the calculations of the depth-dependency of the CO photodissociation rate has not yet been performed. 
Wind acceleration affects the density profile and shielding efficiency in the inner envelope. This will affect the lowest density envelopes where the external UV photons penetrate into the acceleration region.
This remains to be improved in future models.
The \cite{Draine78} radiation field is used to represent the interstellar radiation field.

We used the gas kinetic temperature profile given by \cite{Mamon88} for all models,
\begin{equation}
T_{\rm kin}(r) = 14.6 \: (\frac{r_0}{r})^{\beta} \:\:\:\: [\rm K],
\label{Temp}
\end{equation}
where $r_0$ = 9 $\times$ 10$^{16}$\,cm and $\beta$\,=\,0.72 for $r<r_0$ and $\beta$\,=\,0.54 for $r>r_0$. We assumed a minimum temperature of 10\,K in the outer CSE which is also the minimum limit in the calculations of the chemical reaction rates.

In the chemical models, the dust absorption is assumed to be independent of the wavelength in the spectral region of interest. We also assumed that the dust absorption dominates the dust extinction. Therefore, we used the dust extinction at $1000 \: \r{A}$ as given by \cite{Morris83},
\begin{equation}
\tau_{\rm dust} (r, 1000 \r{A}) = \frac{4.65 \times 2 \times N_{\rm H_2}(r)}{1.87 \times 10^{21}},
\end{equation}
where $N_{\rm H_2}$ is the H$_2$ column density to infinity in cm$^{-2}$.

We adopted a gas-to-dust mass ratio of 200, a grain radius of 0.1\,$\mu$m, and a dust grain density $\rho$\,=\,3.5\,g\,cm$^{-3}$ for all models. Although the dust grain density in C-type CSEs is expected to be lower than the adopted value here, this assumption is not expected to affect spatial extent of the CO and HCN abundance distributions.

\section{Results of the chemical models}\label{Results and Discussion}

\subsection{CO isotopologues}

\begin{figure*}
\centering
\begin{minipage}{.5\textwidth}
  \centering
    \includegraphics[width=90mm]{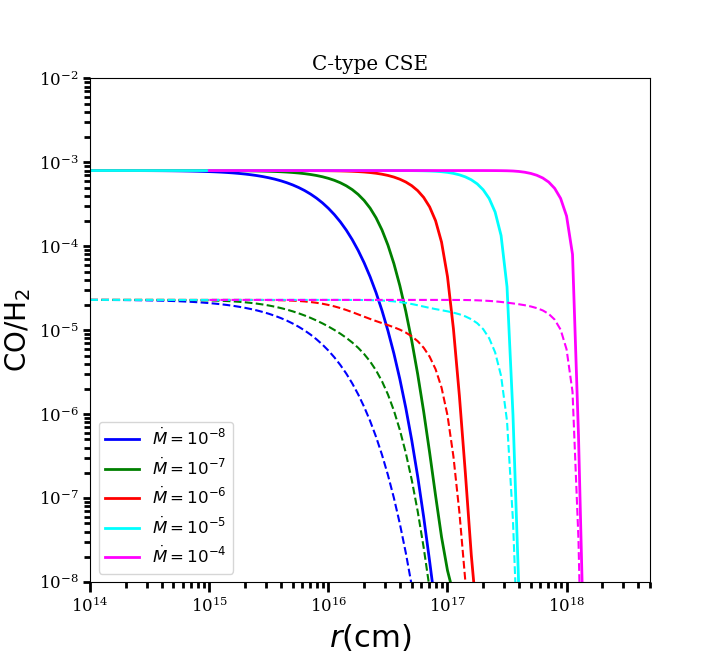}
\end{minipage}%
\begin{minipage}{.5\textwidth}
  \centering
  \includegraphics[width=90mm]{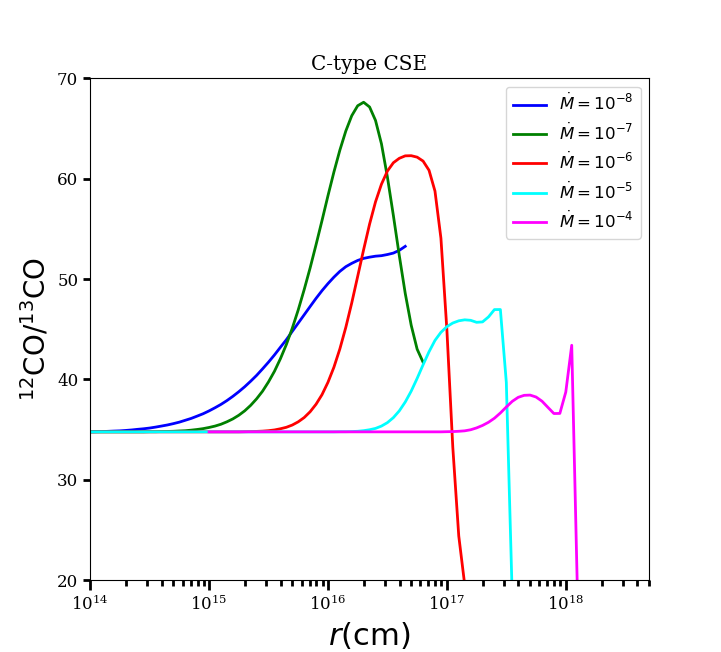}
\end{minipage}
\caption[]{\label{COdist-Ctype} Left:  $\rm ^{12}CO$ (solid lines) and $\rm ^{13}CO$ (dashed lines) abundance profiles for C-type CSE models with different mass-loss rates (in units of $M_\odot\, \rm yr^{-1}$). Right:  $\rm ^{12}CO/^{13}CO$ abundance ratios corresponding to the abundances profiles in the left panel.}
\end{figure*}

In the first set of analyses, we investigated the $^{12}$CO and $^{13}$CO abundance distributions and their corresponding $^{12}$CO/$^{13}$CO ratio in the reference models for the CSEs of C-type and O-type.
Since the results show that the abundance-radius relationships are similar, albeit having scaled shapes, for C- and O-type CSEs, we only present the results for the C-type CSEs here, in Fig.~\ref{COdist-Ctype}, while those for the O-type CSEs are given in Fig.~\ref{COdist-Mtype}.
As shown, the $^{13}$CO abundance generally starts to drop at smaller radii and more gradually than that of $^{12}$CO for all models due to a lower $^{13}$CO shielding efficiency.
At higher mass-loss rates the abundances of both isotopologues drop off more sharply due to higher shielding efficiencies. Deviations of the abundance ratio from the assumed initial value at the inner edge of the model occur at smaller radii with decreasing mass-loss rate.
Table \ref{RCO-C} lists the radii at which the two CO isotopologue abundances drop to half of their initial values and the radii where the $^{12}$CO/$^{13}$CO abundance ratios reach the maximum deviation from its initial value of 34 for C-type CSEs. The molecular initial abundances of two isotopologues and the competition between the photodissociation efficiencies and the fractionation reaction determines the magnitude of the deviation and the radius where it occures.
For stars with $\dot{M}<10^{-5}$ $M_{\odot}$\, yr$^{-1}$, the CO isotopologue ratio starts deviation at a smaller radius than where the $^{12}$CO abundance drops from its initial value, meaning that photodissociation of $^{13}$CO mainly controls the $^{12}$CO/$^{13}$CO ratio.
In the outer regions, as the temperature decreases, the fractionation reaction favours creation of $^{13}$CO from $^{12}$CO resulting in the decrease of the $^{12}$CO/$^{13}$CO ratio for all models.

\begin{table}[t]
  \centering
  \setlength{\tabcolsep}{4.5pt}
    \caption{\tiny Results of the chemical model for CO in C-type CSEs.}
  \begin{tabular}{@{} clclc@{}}
\hline\hline
$\dot{M}$ & $r_{1/2}(^{12}$CO)$^a$ & $r_{1/2}(^{13}$CO)$^a$ & $r_{\rm dev_{\rm max}}^b$ & dev$_{\rm max}^c$ \\
 $(M_{\odot}$ yr$^{-1})$  &  (cm) &  (cm) &  (cm) & \\
\hline
$10^{-8}$ &  $7.6 \times 10^{15}$ & $5.4 \times 10^{15}$ & $4.5 \times 10^{16}$ & $53\%$ \\
$10^{-7}$ & $1.8 \times 10^{16}$ & $1.0 \times 10^{16}$ & $2.0 \times 10^{16}$ & $94\%$ \\
$10^{-6}$ & $6.1 \times 10^{16}$ & $3.3 \times 10^{16}$ & $5.0 \times 10^{16}$ & $79\%$ \\
$10^{-5}$ & $2.1 \times 10^{17}$ & $1.7 \times 10^{17}$ & $2.8 \times 10^{17}$ & $35\%$ \\
$10^{-4}$ & $8.6 \times 10^{17}$ & $8.1 \times 10^{17}$ & $1.1 \times 10^{18}$ & $25\%$\\
\hline
 \end{tabular}
\tablefoot{($^a$) Radius at which the CO isotopologue abundances have reached half of their initial values; ($^b$) The radius at which the $^{12}$CO/$^{13}$CO abundance ratio has its maximum deviation from its initial value; ($^c$) The maximum deviation of the $^{12}$CO/$^{13}$CO abundance ratio from its initial value.}
 \label{RCO-C}
\end{table}


\subsubsection{The impact of the ISRF intensity on the CO isotopologue ratio}

\begin{figure*}
\centering
\begin{minipage}{.5\textwidth}
  \centering
  \includegraphics[width=92mm]{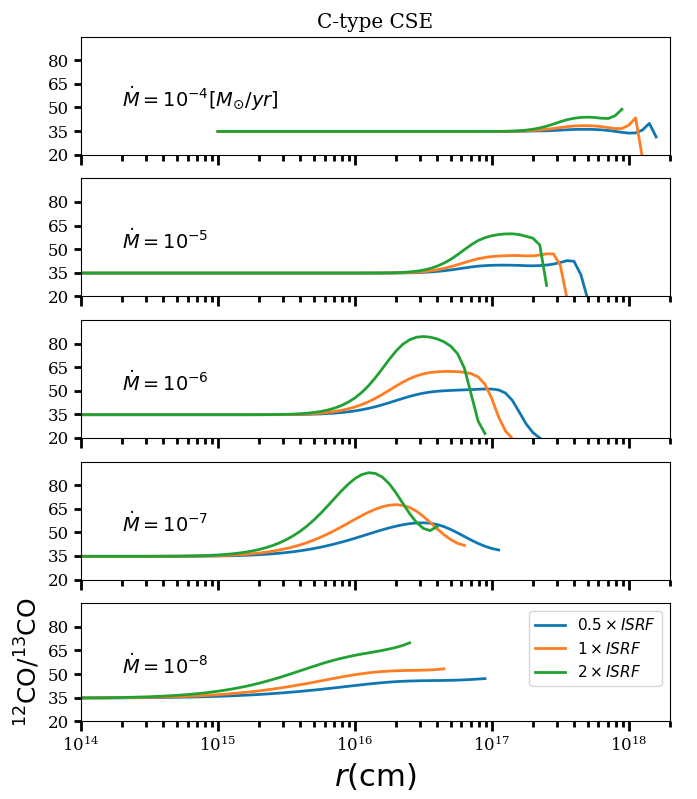}
\end{minipage}%
\begin{minipage}{.5\textwidth}
  \centering
  \includegraphics[width=92mm]{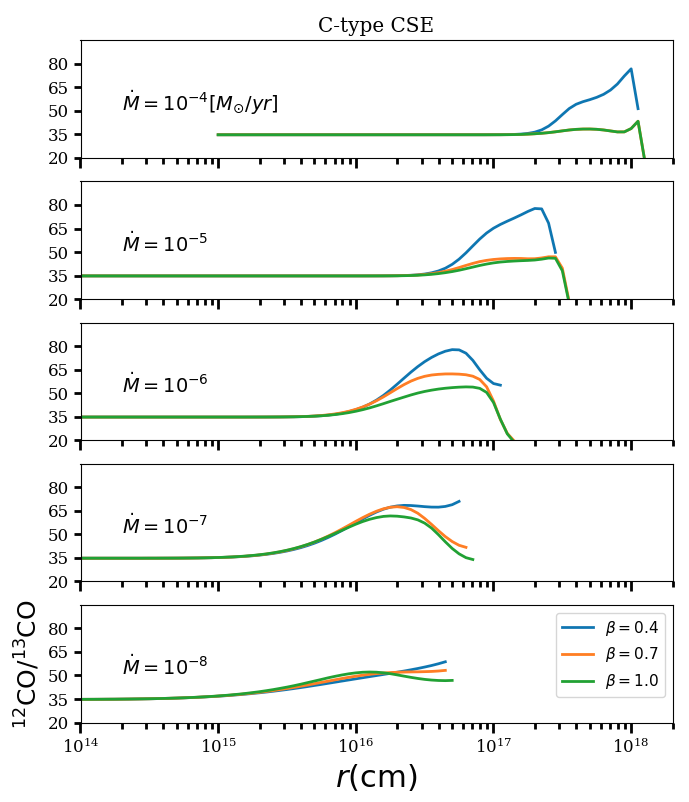}
\end{minipage}
\caption[]{\label{Chemi-var-C} $\rm ^{12}CO/^{13}CO$ abundance ratios through C-type CSE models with different mass-loss rates under variations of the ISRF intensity (left) and the kinetic temperature profile (right).}
\end{figure*}

In photochemical models of CSEs, the standard \cite{Draine78} radiation field is usually assumed to penetrate the CSE. This radiation field is based on measurements in the solar vicinity. Therefore, it may not be a good representative of the radiation field that influences the CSE chemistry of AGB stars which are located, for instance, in globular clusters, above the Galactic plane, or in the vicinity of star-forming clusters \citep[e.g.,][]{McDonaldZ15}.
\cite{Saberi19} showed that variations of the interstellar radiation field (ISRF) intensity have a considerable impact on the CO photodissociation rate.
In order to estimate how much the ISRF intensity affects the CO isotopologue ratio in CSEs of AGB stars with different mass-loss rates, we scaled the ISRF by factors of 0.5 and 2.
The results are presented in the left panel of Fig.~\ref{Chemi-var-C}. 

For a stronger ISRF and stars with lower mass-loss rates, the deviation of the CO isotopologue abundance ratio from atmospheric value starts at smaller radii due to the deeper penetration of the UV photons. 
The largest impact of the variation of the ISRF on the CO isotopologue ratio is seen for models with $\dot{M}=10^{-6}$ and $10^{-7}$ $M_{\odot}$ yr$^{-1}$.
For stars with higher mass-loss rates, almost all the relevant photons are absorbed in the outer part of the CSE, and for the ones with lower mass-loss rates, the shielding efficiency is anyhow low due to the low gas density.

We note that in a clumpy CSE, the ISRF can penetrate deeper into the CSE depending on the degree of the clumpiness. 
On the other hand, clumpiness will increase the local density (over that of a smooth CSE formed by the same mass-loss rate) and hence increase the shielding efficiency.
To calculate the impact of clumpiness on the CO isotopologue distribution require introduction of several parameters such as, the size of clumps and their location and distribution which are hard to observationally constrain. Such a study is beyond the scope of this paper.

\subsubsection{The impact of the gas kinetic temperature on the CO isotopologue ratio}

The gas kinetic temperature in an AGB CSE can be reasonably approximated by a power-law (Eq. \ref{Temp}) with $0.4<\beta<1$ \citep[e.g.,][]{DeBeck12, Danilovich14,Khouri14, Maercker16, Ramos18}.
The temperature affects the fractionation reaction rate and the photodissociation rate (e.g., increasing temperature leads to a lower CO shielding efficiency which is explained in detail in \cite{Saberi19}). 
To investigate to what extent a change in the gas temperature affects the CO isotopologue ratio, we considered values of $\beta$\,=\,0.4, 0.7, and 1 in Eq.~\ref{Temp}, and assumed the temperature and radius at the inner envelope to be $T_0 = 2000$ K and $r_0 = 10^{14}$ cm. The results are shown in the right panel of Fig.~\ref{Chemi-var-C}. 
In particular for $\beta=0.4$ which results in a warmer CSE, the CO isotopologue ratio in the outer layers can be strongly affected.
This is because the forward and backward fractionation reaction rates are equal at high temperatures and consequently the fractionation reaction does not change the isotopologue ratio. 
Only the isotopologue-selective photodissociation will have a significant influence. Since $^{12}$CO is more efficiently shielded than $^{13}$CO, we find an increase in the $^{12}$CO/$^{13}$CO. This effect is especially clear for high mass-loss rates, where the shielding efficiencies are higher.

\subsection{HCN isotopologues}

The photodissociation of HCN does not occur in lines. Instead photons in a broad UV range can lead to dissociation of the HCN molecule, and the photodissociation cross section is therefor continuous in wavelength \cite[e.g.,][]{vanDishoeck11, Saberi17}. Therefore, the H$^{12}$CN/H$^{13}$CN ratio is not expected to be affected by selective photodissociation.

\begin{figure*}
\centering
\begin{minipage}{.5\textwidth}
  \centering
    \includegraphics[width=100mm]{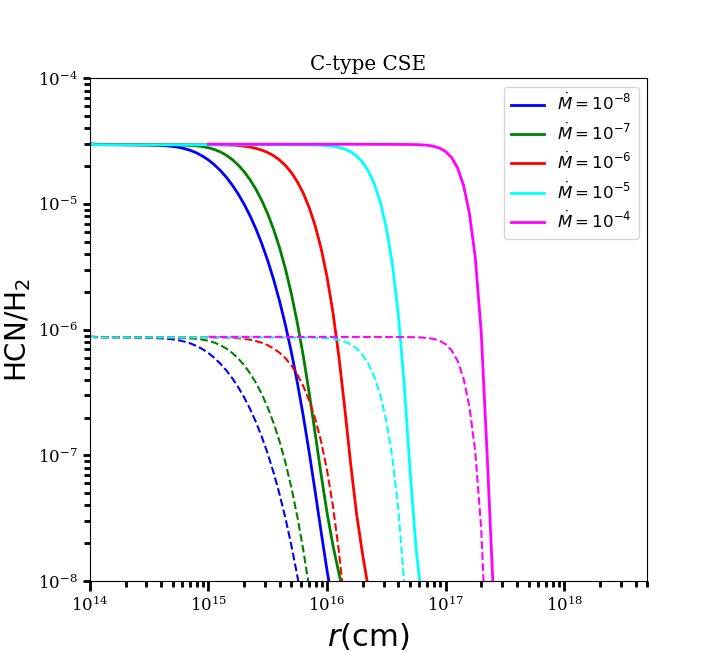}
\end{minipage}%
\begin{minipage}{.5\textwidth}
  \centering
    \includegraphics[width=100mm]{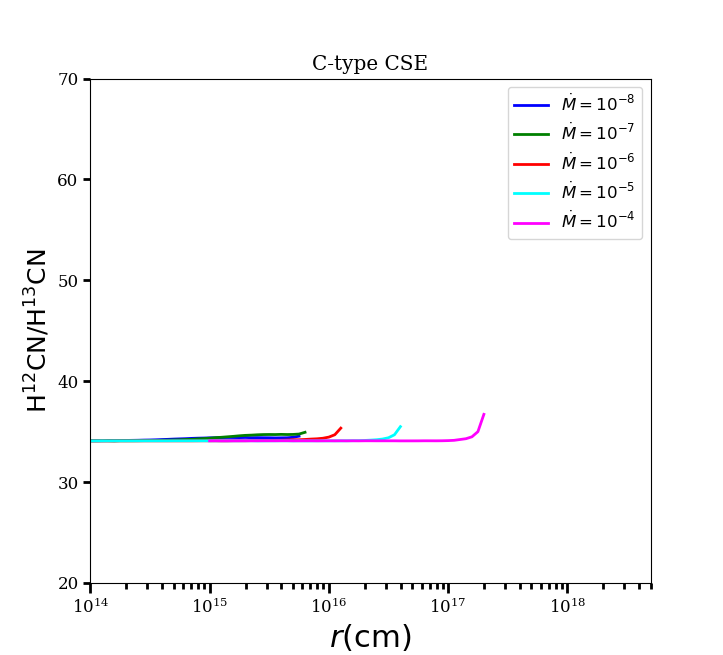}
\end{minipage}
\caption[]{\label{HCN-C} Left:  $\rm H^{12}CN$ (solid lines) and $\rm H^{13}CN$ (dashed lines) abundance profiles for C-type CSE models with different mass-loss rates (in units of $M_\odot\, \rm yr^{-1}$). Right:  $\rm H^{12}CN/H^{13}CN$ abundance ratios corresponding to the abundances profiles in the left panel.}
\end{figure*}

Figure~\ref{HCN-C} presents the H$^{12}$CN and H$^{13}$CN abundance distributions, and their isotopologue ratio, for C-type CSEs. As expected, variations of the HCN isotopologue ratio are negligible for all models. The results for O-type CSEs are presented in Fig.~\ref{HCN-O}.

\section{Radiative transfer modeling}

\begin{table}[t]
  \centering
  \setlength{\tabcolsep}{4.5pt}
    \caption{\tiny Model assumptions for reference models with different mass-loss rates used in the radiative transfer modeling.}
  \begin{tabular}{@{} ccccccc@{}}
\hline\hline
$R_{\star}$ & $L_{\star}$ & $T_{\star}$ &  $V_{\rm exp}$ & $T_{\rm dust}$$^1$ & $R_{\rm in} $\\
\:
 (cm)  & ($L_{\odot}$) & (K)  & (km s$^{-1}$) & (K) &  (cm) \\
\hline
$3.5\times10^{13}$ & 9000 & 2500 & 15  &  $T_{\rm in} (R_{\rm in}/r)^{0.4}$ & $1 \times 10^{14}$\\
\hline
 \end{tabular}
\tablefoot{1. $T_{\rm in}$ is the dust condensation temperature and is assumed to be 1200 and 1500 K for C-type and O-type CSEs, respectively.}
 \label{SModel}
\end{table}

\subsection{Modeling the molecular line emission}

For the radiative transfer (RT) analysis, we used a non-local thermodynamic equilibrium (non-LTE) radiative transfer code based on the Monte Carlo method \citep[mcp; e.g.,][]{Bernes79, Schoier01}. In the case of HCN, we used a RT code based on the accelerated lambda iteration formalism for models with mass-loss rates $>$\,10$^{-6}$\,$M_\odot$\,yr$^{-1}$ to handle the high optical depths in the H$^{12}$CN lines in the C-type CSEs \citep[e.g.,][]{maercker08}.

The CO excitation analysis includes 82 rotational energy levels, up to $J$\,=\,40 within in the ground ($\nu=0$) and first vibrationally excited ($\nu=1$) states. Radiative transitions within and between the vibrational states are taken into account.
The collisional rate coefficients between CO and para- and ortho-H$_2$ are taken from \cite{Yang10}. 
An ortho-to-para-H$_2$ ratio of 3 was adopted for weighting the ortho-to-para-H$_2$ coefficients together. The rates cover 25 temperatures between 2 to 3000\,K. The same range of transitions and rates is used for both CO isotopologues.

For the HCN excitation analysis, we included 126 rotational energy levels, up to $J$\,=\,29 within the vibrational ground state and first excited states for each of the three vibrational modes $\nu_1$, $\nu_2$, and $\nu_3$. Hyperfine splitting is included for the $J$\,=\,1 state in the ground vibrational state and for the lowest rotational state in each excited vibrational state. 
The collisional rate coefficients between HCN and H$_2$ are scaled by a factor of 1.363 from the collisional rates between HCN and He calculated by \cite{Dumouchel10}. We used results for temperatures between 5 and 500\,K. The same range of transitions and rates is used for both HCN isotopologues.

Similar to our chemical models, we assume a CSE expanding with constant velocity, which is formed by a constant and isotropic mass-loss. 
The gas kinetic temperature is the same as in our chemical models, Eq.~\ref{Temp}. 
The stellar and CSE parameters of the reference models are presented in Table~\ref{SModel}.
The CO and HCN isotopologue abundance distributions are taken from our chemical models as presented in the left panels of Figs.~\ref{COdist-Ctype} and \ref{HCN-C}. 

We adopted amorphous carbon dust \citep{Suh00} and amorphous silicate dust \citep{Justtanont92} for the C-type and O-type CSEs, respectively.
The dust condensation temperature is assumed to be 1200 and 1500\,K for carbon and silicate dust, respectively \citep{DeBeck10}, and the dust temperature radial profile is given in Table~\ref{SModel}.
We assumed that the dust optical depth is 0.013 for the model with $\dot{M}=10^{-7} M_{\odot}\,$yr$^{-1}$ at 10\,$\mu$m ($\tau_{10}$), corresponding to the U Hya model of \cite{Danilovich15}, and then scaled it linearly with the mass-loss rate.

The reference stars are assumed to be at distances of 200, 300, 500, 840, and 3100\,pc for stars with mass-loss rates $10^{-8}, 10^{-7}$,..., and $10^{-4}$ $M_{\odot}$ yr$^{-1}$, respectively, in order to cover the entire emitting region with the beam of a 12\,m telescope, e.g. the APEX telescope.
 The half-power beam width used to extract the spectra is given by
 \begin{equation}
\theta = 7.8 \frac{800}{\nu[GHz]} \,\, [^{\prime\prime}],
\end{equation}
 where $\nu$ is the line frequency.

\subsection{CO isotopologues}

\begin{figure*}
\centering
\begin{minipage}{.5\textwidth}
  \centering
  \includegraphics[width=100mm]{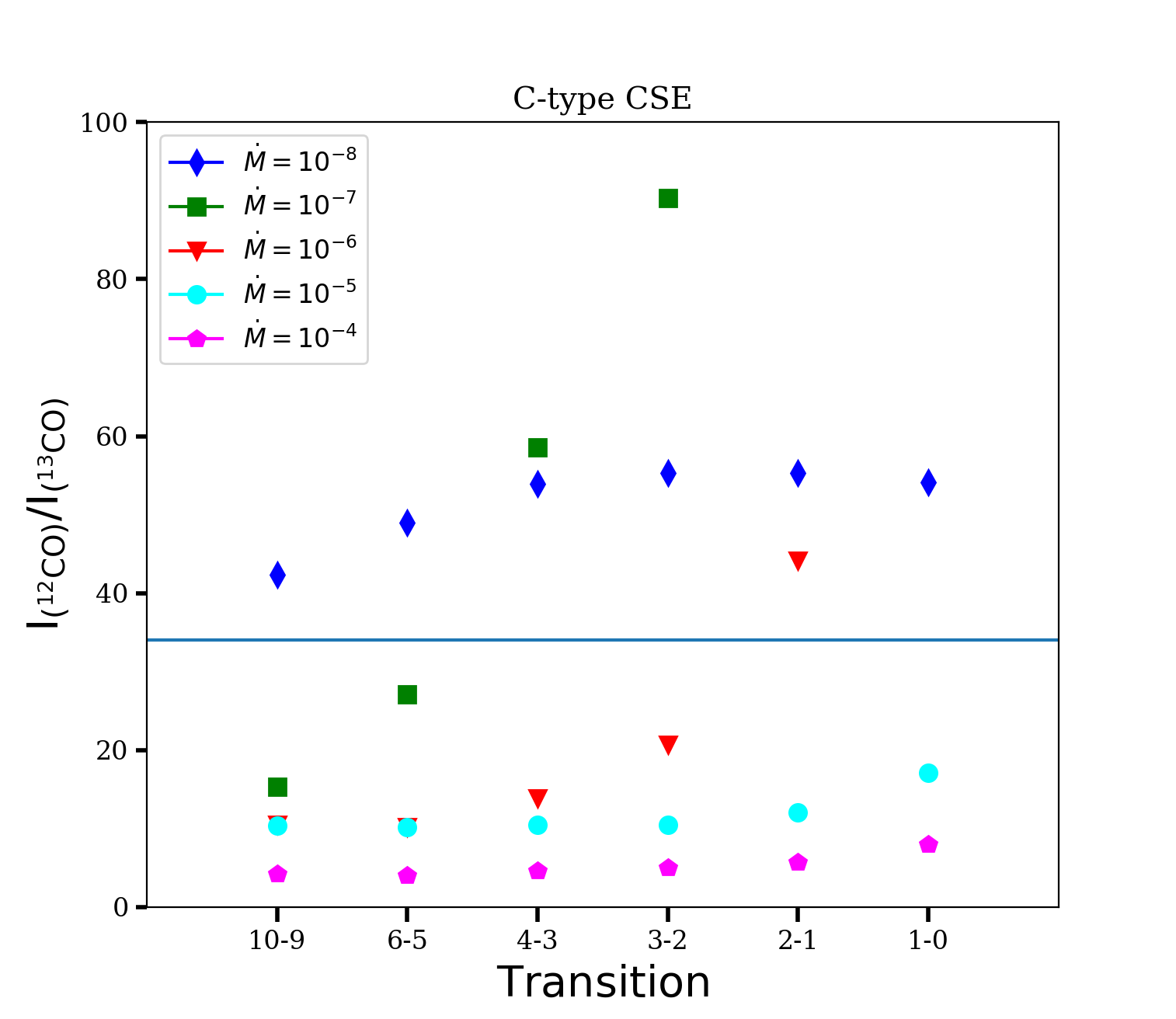}
\end{minipage}%
\begin{minipage}{.5\textwidth}
  \centering
  \includegraphics[width=100mm]{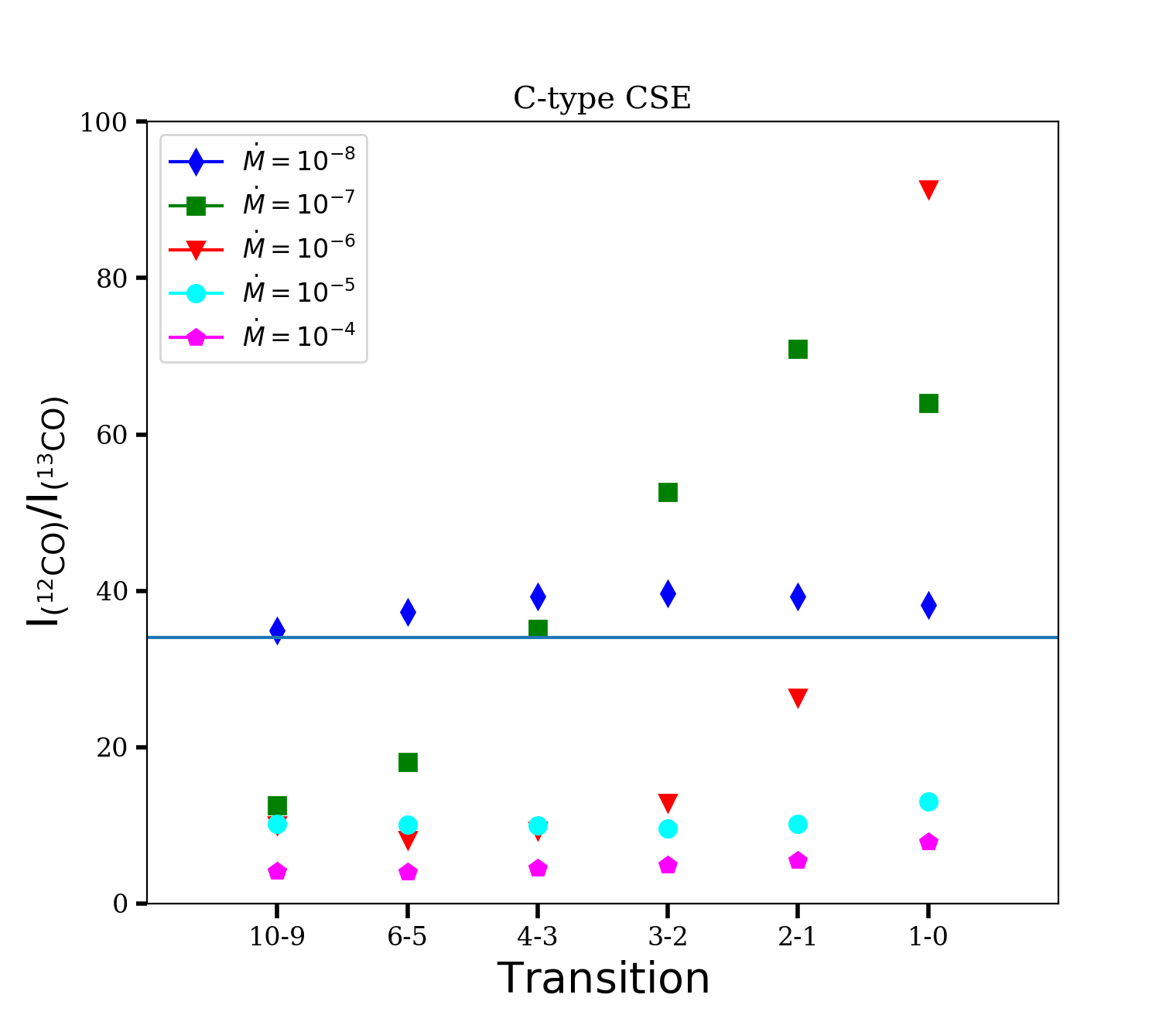}
\end{minipage}
\caption[]{\label{COinyensityratio-C} Line intensity ratios $I_{\rm ^{12}CO}/I_{\rm ^{13}CO}$ from RT reference models with different mass-loss rates and with (left) and without (right) considering chemical effects in determining the abundances profiles ((points exceeding values higher than 100 are not plotted)). The solid blue lines show the adopted initial isotopologue abundance ratio of 34 for C-type CSEs.}
\end{figure*}

We derived the integrated line intensity ratio $I_{\rm ^{12}CO}/I_{\rm ^{13}CO}$ for six rotational transitions $J$\,=\,\mbox{1--0}, \mbox{2--1}, \mbox{3--2}, \mbox{4--3}, \mbox{6--5}, and \mbox{10--9} in the ground-state in order to reasonably cover emission from the entire CO envelope ($I$\,=\,$\int T_{\rm mb} {\rm d}\upsilon$, where $T_{\rm mb}$ is the main beam brightness temperature). Again, since the general trends for all models are very similar for C- and O-type CSEs, we present the results of C-type CSEs here, and those of the O-type CSEs in Sect.~\ref{M-type-RT}.

The results for the C-type CSEs are shown in the left panel of Fig.~\ref{COinyensityratio-C}. There are large variations in the line intensity ratios of different rotational transitions for stars with intermediate mass-loss rates (about $\dot{M}\approx10^{-7}$ to $\approx10^{-6}$\,$M_{\odot}$\,yr$^{-1}$), while for stars with lower and higher mass-loss rates the ratios vary little with transition. This is expected from the ratios of the abundance distributions shown in Fig. \ref{COdist-Ctype}, and in particular from the excitation properties of the CO molecule.
The CO isotopologues go from being predominantly radiatively excited at lower densities to being collisionally excited at higher densities. This transition occurs for intermediate mass-loss rates ($\sim 10^{-7}-10^{-6} M_{\odot} \rm yr^{-1}$), which are slightly different for the two isotopologues, see App.\ref{TauCO} and \cite{Khouri14}.
In particular, the lower-$J$ lines are strongly affected, and the line intensity ratios can differ significantly from the isotopologue abundance ratio. The difference in excitation behaviour of the CO isotopologue rotational transitions in this range of mass-loss rate are illustrated in App.~\ref{TauCO}. Although the line intensity ratios for stars with high mass-loss rates vary little with transition, the ratio is shifted to a lower value than the isotopologue abundance ratio. This is because of saturation in the $^{12}$CO lines, due to high optical depths, that leads to a lower $^{12}$CO/$^{13}$CO line intensity ratio.

In order to evaluate the sensitivity of the CO isotopologue model line intensity ratios to the main input parameters, we varied several of them over reasonable ranges. First, we investigate how much our derived abundance distributions from the detailed chemical model affect the line intensity ratios.
We ran a new set of RT models by adopting the $^{12}$CO abundance distributions from the chemical analysis and then scale it down by factors of 34 and 13 to derive the $^{13}$CO abundance distributions for C-type and O-type CSEs, respectively. The results for C-type CSEs are presented in the right panel of Fig.~\ref{COinyensityratio-C}, and variations relative to the reference models in percentage are listed under the "No chemistry" label in Table~\ref{variablesC}. The results show that involving chemistry in deriving the molecular abundances can indeed significantly affect the derived line intensity ratios for stars with low mass-loss rates ($\leq10^{-6} M_{\odot}$yr$^{-1}$). Not including the chemical abundance distributions leads to an overestimate of the $^{12}$CO/$^{13}$CO ratio even when performing a RT analysis. The effect is less for the higher-$J$ transitions.

We also examined the effects of variations of $T_{\star}$, $L_{\star}$, and $R_{\rm in}$ by $\pm\,20\,\%$ on the RT results. The results are listed in Table~\ref{variablesC} for C-type CSEs. 
Generally, the models with $10^{-7} \leq \dot{M} \leq 10^{-6} M_{\odot}$ yr$^{-1}$ are the ones most sensitive to variations in input parameters. Varying the stellar parameters $T_{\star}$ and $L_{\star}$ changes the line intensity ratios by less than 17\,$\%$ for all models, and much less than this for the majority of the models.
Variation of the inner radius $R_{\rm in}$ affects mainly the low-$J$ transitions of the $10^{-6} M_{\odot}$ yr$^{-1}$ model. This is due to the sensitivity of the excitation of the CO isotopologues to the gas density and the intensity of the dust radiation field at this mass-loss rate as discussed above.

\subsection{HCN isotopologues}

\begin{figure}[t]
  \centering
  \includegraphics[width=90mm]{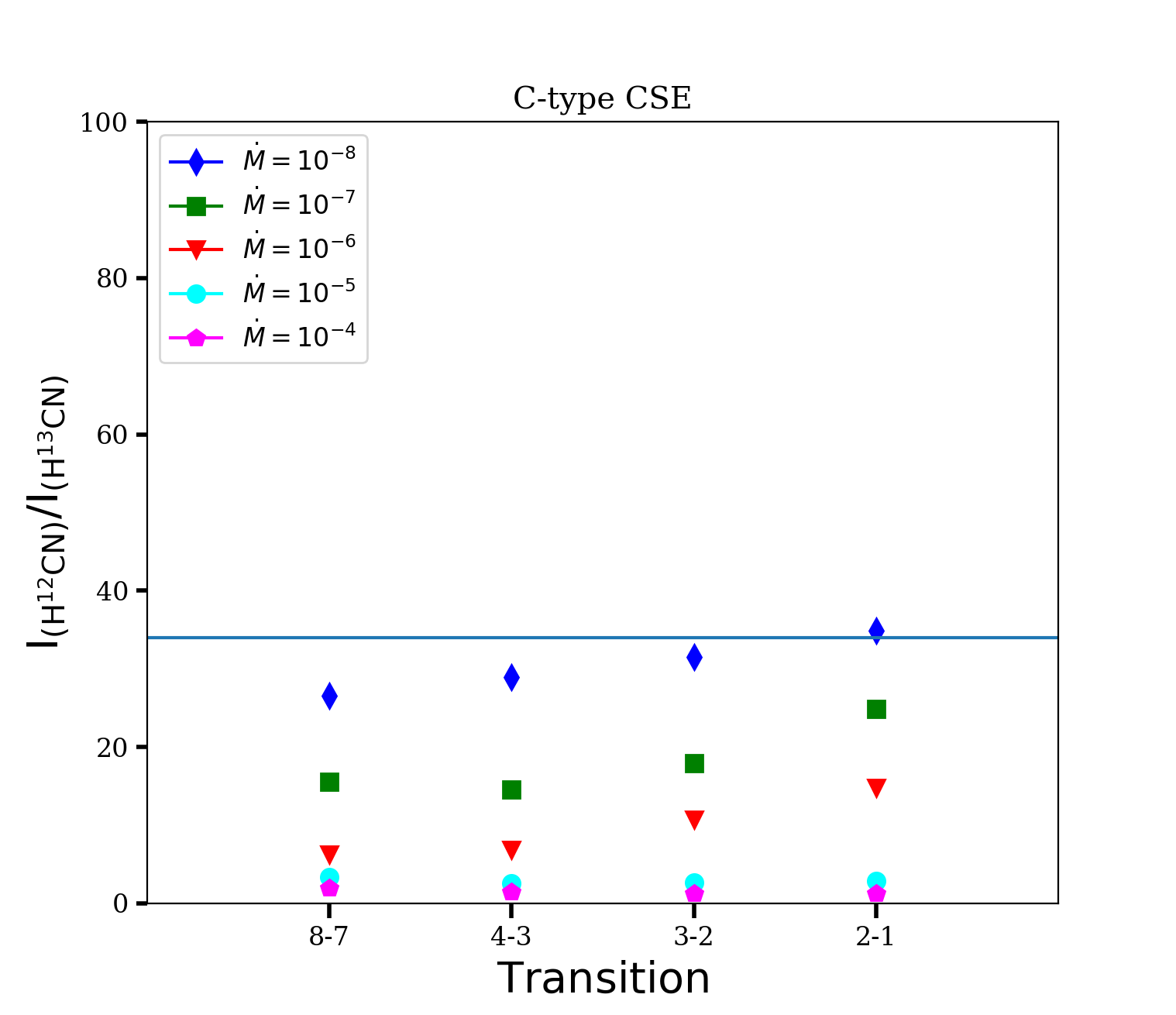}
  \caption[]{
    line intensity ratio $I_{\rm H^{12}CN}/I_{\rm H^{13}CN}$ from RT modeling for the reference models with different mass-loss rates. The solid blue line shows the initial adopted abundance ratio of 34 for C-type CSEs.}
\label{HCN-RT}  
\end{figure}

In the case of HCN, we calculated the integrated line intensity for four rotational transitions $J$\,=\,\mbox{2--1}, \mbox{3--2}, \mbox{4--3}, and \mbox{8--7} to reasonably trace the entire HCN envelope.
The $I_{\rm H^{12}CN}/I_{\rm H^{13}CN}$ ratios extracted from the RT models are shown in Fig.~\ref{HCN-RT} for C-type CSEs. Variations between the line intensity ratios of different HCN rotational transitions are significantly smaller than those obtained for CO (especially for the intermediate-mass-loss-rate sources). The line intensity ratios of the high-mass-loss-rate models are shifted to even lower values than in the case of CO due to the higher optical depths in the HCN lines.

\section{Summary}\label{Summary}

We examined variations of the $^{12}$CO/$^{13}$CO and H$^{12}$CN/H$^{13}$CN abundance ratios in AGB CSEs of C-type and O-type over a wide range of model parameters. The abundance distributions derived from our chemical model are used in RT models to provide an excitation analysis. Here, we highlight the main results and their implications.

\subsection{CO}

Our detailed chemical analysis demonstrates that the circumstellar $^{12}$CO/$^{13}$CO abundance ratio can vary significantly from the adopted atmospheric value through the CSEs of both C-type and O-type AGB CSEs. 
Competition between the isotopologue-selective UV photodissociation process and the chemical fractionation reaction determines the magnitude and starting radius of the variations. The efficiencies of these processes are sensitive to the intensity of the UV radiation in the environment of the CSE and the CSE properties such as the gas density and kinetic temperature. Therefore, our results indicate that a detailed chemical analysis for each individual star is needed to determine the CO isotopologue distributions and their corresponding ratio with high accuracy.

Our excitation analysis shows that the line intensity ratio $I_{\rm ^{12}CO}/I_{\rm ^{13}CO}$ of several rotational transitions can deviate significantly from the adopted initial CO isotopologue abundance ratio, in particular for stars with mass-loss rates in the range 10$^{-7}$ to 10$^{-6}$\,$M_{\odot}$\,yr$^{-1}$. This is due to a combination of chemical and excitation effects.
For the higher-mass-loss rate stars, the differences in the line intensity ratios of different transitions are negligible, but the ratios are shifted to lower values than the isotopologue ratio due to optical depth effects.

Our predictions can be tested observationally by using, for example, ALMA. Such observations are underway \citep[e.g.][in press]{Ramstedt20}, but the difficulty lies in converting observed brightness distributions into abundance distributions that can be directly compared to the results of the models.

\subsection{HCN}

The H$^{12}$CN/H$^{13}$CN isotopologue abundance ratio is quite stable throughout the CSEs for all tested models for both C-type and O-type CSEs. The integrated line intensity ratio H$^{12}$CN/H$^{13}$CN differs little between the rotational transitions. The corrections for optical depth effects are substantially larger for HCN than for CO in C-type CSEs where the HCN abundance is high.

\subsection{Consequences for previously estimated circumstellar  \texorpdfstring{$^{12}$CO/$^{13}$CO \,} rratios}

Circumstellar CO isotopologue ratios have often been used to estimate stellar C isotope ratios \citep[e.g.,][]{Schoier00, Milam09, Ramstedt14}. The results are compared with those of stellar nucleosynthesis models and constraints on these are inferred. Therefore, it is important to emphasize that the present study shows that some of the assumptions made in the earlier studies are not correct. However, it remains to estimate how severe the effect is. Based on line intensity ratios of individual transitions, the former studies are likely to have overestimated the $^{12}$CO/$^{13}$CO ratio, and by inference the $^{12}$C/$^{13}$C ratio. However, the earlier studies are based on multi-transition observations and individual models (e.g., determinations of the kinetic temperature radial profile of the CSE) for each source making the effect of the incorrect assumptions, in all likelihood, smaller. Further, the effect depends (strongly) on the mass-loss rate, being much smaller for the lower and higher mass-loss rate sources, and on the transition, being smaller for the higher-$J$ transitions. We conclude that previous studies have, if anything, overestimated the $^{12}$CO/$^{13}$CO ratio, but the magnitude of the effect is difficult to estimate.

In principle, the use of circumstellar HCN isotopologue line emission would be a more reliable tracer of the photospheric $^{12}$C/$^{13}$C ratio. However, there are two things to note. First, in the C-type CSEs the HCN lines are much more saturated than the CO lines and a proper radiative transfer analysis taking optical depth effects into account is required. Second, in O-type CSEs the HCN abundance is low, and it becomes difficult to detect the H$^{13}$CN lines in many sources.

\subsection{A comment on the \texorpdfstring{C$^{17}$O/C$^{18}$O \,} iisotopologues ratio}
The circumstellar C$^{17}$O/C$^{18}$O isotopologue ratio has been used to estimate the stellar $^{17}$O/$^{18}$O isotope ratio \citep[e.g.,][]{denutte17}. However, in this case we expect only a small difference in behaviour between the C$^{17}$O and C$^{18}$O isotopologoues. The reason is that differences in photodissociation and chemical fractionation efficiencies are expected to be smaller for these species than for $^{12}$CO and $^{13}$CO due to their more comparable masses.
In addition,  their spectral lines are more optically thin due to their lower abundances.

The isotope ratios $^{16}$O/$^{17}$O and $^{16}$O/$^{18}$O will necessarily involve the $^{12}$C$^{16}$O molecule and are thus affected by the same problem as discussed in this paper (use of the rarer isotopologues $^{13}$C$^{17}$O  and $^{13}$C$^{18}$O is not possible since they produce too weak line emission in CSEs to be detected).

\begin{table*}[t]
  \centering
  \setlength{\tabcolsep}{5pt}
  \renewcommand{\arraystretch}{1.5}
    \caption{\tiny Change in the $^{12}$CO/$^{13}$CO line intensity ratio, relative to the reference models, when varying different input parameters in the RT modeling for C-type CSEs.}
  \begin{tabular}{@{}|c|c|cc|cc|cc|cc|cc|cc|@{}}
\hline
$\dot{M}$ & Variable &  \multicolumn{12}{c}{Variation of $I_{^{12}\rm CO}/I_{^{13}\rm CO}$ relative to reference models} \vline \\
\hline
    &   & \multicolumn{2}{c} {${J=1\rightarrow0}$} & \multicolumn{2}{c}{${2\rightarrow1}$}  &  \multicolumn{2}{c}{${3\rightarrow2}$} &  \multicolumn{2}{c}{${4\rightarrow3}$}  & \multicolumn{2}{c} {${6\rightarrow5}$}  & \multicolumn{2}{c} {${10\rightarrow9}$} \vline \\
  & & \multicolumn{2}{c} {$E_{\rm up}$=5 (K)} & \multicolumn{2}{c}{$E_{\rm up}$=17}  & \multicolumn{2}{c}{$E_{\rm up}$=33} & \multicolumn{2}{c}{$E_{\rm up}$=55} & \multicolumn{2}{c}{$E_{\rm up}$=116} & \multicolumn{2}{c}{$E_{\rm up}$=304} \vline \\
\hline
& $T_\star$ & $-20\%$ &$ +20\%$ & $-20\%$ &$ +20\%$& $-20\%$ &$ +20\%$& $-20\%$ &$ +20\%$& $-20\%$ &$ +20\%$ & $-20\%$ &$ +20\%$\\
\hline
$10^{- 8 }$&& 0 \%& -1 \%& 0 \%& -1 \%& 1 \%& -2 \%& 2 \%& -2 \%& 4 \%& -4 \%& 4 \%& -3 \%\\
$10^{- 7 }$&& -7 \%& 7 \%& 7 \%& -7 \%& 17 \%& -12 \%& 20 \%& -13 \%& 13 \%& -8 \%& -5 \%& 5 \%\\
$10^{- 6 }$&& 8 \%& -4 \%& 4 \%& -2 \%& 2 \%& 0 \%& -1 \%& 0 \%& -3 \%& 2 \%& -3 \%& 2 \%\\
$10^{- 5 }$&& 0 \%& 0 \%& -1 \%& 0 \%& 0 \%& 0 \%& 0 \%& 0 \%& 0 \%& 0 \%& 0 \%& 0 \%\\
$10^{- 4 }$&& 2 \%& 0 \%& 0 \%& -2 \%& -1 \%& -2 \%& -1 \%& 0 \%& 0 \%& 0 \%& 0 \%& 0 \%\\
\hline
& $L_\star$ & $-20\%$ &$ +20\%$ & $-20\%$ &$ +20\%$& $-20\%$ &$ +20\%$& $-20\%$ &$ +20\%$& $-20\%$ &$ +20\%$ & $-20\%$ &$ +20\%$\\
\hline
$10^{- 8 }$&& 1 \%& 0 \%& 1 \%& 0 \%& 0 \%& 0 \%& -1 \%& 1 \%& -2 \%& 2 \%& -3 \%& 2 \%\\
$10^{- 7 }$&& 5 \%& -2 \%& -3 \%& 3 \%& -6 \%& 7 \%& -8 \%& 8 \%& -5 \%& 4 \%& 1 \%& -3 \%\\
$10^{- 6 }$&& -3 \%& 3 \%& 0 \%& 2 \%& 1 \%& 0 \%& 1 \%& -1 \%& 1 \%& -1 \%& 1 \%& -2 \%\\
$10^{- 5 }$&& -1 \%& -1 \%& -1 \%& -1 \%& 0 \%& 1 \%& 0 \%& 1 \%& 0 \%& 0 \%& 0 \%& 1 \%\\
$10^{- 4 }$&& 6 \%& 1 \%& 1 \%& 0 \%& 0 \%& 0 \%& 0 \%& 0 \%& 0 \%& 0 \%& 0 \%& 0 \%\\
\hline
 &  $R_{in}$ & $-20\%$ &$ +20\%$ & $-20\%$ &$ +20\%$& $-20\%$ &$ +20\%$& $-20\%$ &$ +20\%$& $-20\%$ &$ +20\%$ & $-20\%$ &$ +20\%$\\
\hline
$10^{- 8 }$&& 1 \%& 0 \%& 2 \%& -1 \%& 2 \%& -1 \%& 2 \%& -1 \%& 1 \%& -1 \%& -2 \%& 1 \%\\
$10^{- 7 }$&& 7 \%& -4 \%& 4 \%& -2 \%& -1 \%& 1 \%& -5 \%& 3 \%& -8 \%& 6 \%& -6 \%& 4 \%\\
$10^{- 6 }$&& -9 \%& 7 \%& -5 \%& 6 \%& -2 \%& 3 \%& -1 \%& 1 \%& 2 \%& -1 \%& 4 \%& -3 \%\\
$10^{- 5 }$&& -1 \%& 0 \%& 0 \%& -1 \%& 1 \%& -1 \%& 1 \%& -2 \%& 2 \%& -1 \%& 2 \%& -1 \%\\
$10^{- 4 }$&& 2 \%& -5 \%& -1 \%& 0 \%& -1 \%& 0 \%& 0 \%& 0 \%& 1 \%& 1 \%& 0 \%& 0 \%\\
\hline
& No chemistry & &&&&&&&&&&&\\
\hline
$10^{-8}$   & &-29\%&&  -29\%&&-28\%  &&-27\% && -24\% && -17\% &\\
$10^{-7}$   & &-42\%  &&-42\%  &&-42\%  &&-40\% && -33\% && -18\% & \\
$10^{-6}$   & &-41\%   &&-41\%  &&-38\%  &&-33\% && -20\% && -5\% &\\
$10^{-5}$   & & -24\%  &&-16\%  &&-9\%  &&-4\% && -2\% && -1\% & \\
$10^{-4}$   & &-2 \%  &&-3\%  &&-2\%  &&-2\% && -2\% && -2\% & \\
\hline
 \end{tabular}
 \label{variablesC}
\end{table*}

\begin{acknowledgements}
We would like to thank our referee, Anita Richards, for very good comments and constructive feedback on the manuscript.
This research was supported by the Swedish Research Council (VR). EDB acknowledges financial support from the Swedish National Space Agency. 

\end{acknowledgements}




\bibliographystyle{aa} 
\bibliography{refrences} 


\begin{appendix} 

\section{Chemical network}
\label{s:chemistry}

The initial fractional abundances, relative to H$_2$, of parent species for C-type and O-type CSEs are listed in Table~\ref{ParentS}.
The CO fractional abundance of $5\times10^{-4}$ for O-type AGB CSEs is estimated assuming that $80\%$ of the solar carbon abundance is in the form of CO. This value is higher than the value of $\approx$\,$2\times10^{-4}$ which is commonly used for O-type CSEs, but is more consistent with ALMA-ACA observations of a sample of 21 AGB CSEs (Ramstedt et al.; submitted to A\&A). Likewise, we assumed that $80\%$ of the solar oxygen abundance goes into CO to estimate the CO fractional abundance in C-type CSEs, $8\times10^{-4}$. This value is commonly used in the literature.
The HCN abundances are taken as the median values estimated by \cite{Schoier13} in C-type and O-type CSEs. 
The abundances of the other C- and O-bearing species are taken from the thermal-equilibrium values of IRC+10216 (C-type AGB star) and IK~Tau (M-type AGB star) as given by \cite{Willacy98} and \cite{Duari99}, respectively. The remaining abundances are taken from \cite{McElroy13}.

\begin{table}[t]
  \centering
  \setlength{\tabcolsep}{6pt}
 \caption{Initial fractional abundances, relative to H$_2$, of parent
species for C-type and O-type CSEs.}
\begin{tabular}{c|c|clc|}
\hline\hline
Species & C-type & O-type  \\               
\hline
CO & 8.0$\times 10^{-4}$ & 5.0$\times 10^{-4}$  &  \\
HCN & 3.0$\times 10^{-5}$ & 1.0$\times 10^{-7}$ & \\
CS & 1.3$\times 10^{-5}$ & 1.3$\times 10^{-10}$  &\\
C$_2$H & 5.5$\times 10^{-5}$ & 5.5$\times 10^{-18}$& \\
C$_2$H$_2$ & 1.5$\times 10^{-4}$ & 8.5$\times 10^{-17}$  &\\
SiO & 1.9$\times 10^{-8}$ & 4.3$\times 10^{-5}$  & \\
SiS & 1.3$\times 10^{-6}$ & 1.3$\times 10^{-6}$ & \\
CH$_4$ & 3.5$\times 10^{-6}$ & 1.0$\times 10^{-14}$ & \\
SiC$_2$ & 4.7$\times 10^{-7}$ & 1.0$\times 10^{-14}$ & \\
H$_2$O & 1.0$\times 10^{-7}$ & 1.0$\times 10^{-4}$  & \\
SiH$_4$ & 2.2$\times 10^{-7}$ & 2.2$\times 10^{-7}$ & \\
NH$_3$ & 2.0$\times 10^{-6}$ & 2.0$\times 10^{-6}$ & \\
HCl & 1.0$\times 10^{-7}$ & 1.0$\times 10^{-7}$ &\\
HCP & 2.5$\times 10^{-8}$ & 2.5$\times 10^{-8}$ & \\
HF & 8.0$\times 10^{-9}$ & 8.0$\times 10^{-9}$ & \\
H$_2$S & 4.0$\times 10^{-9}$ & 4.0$\times 10^{-9}$ & \\
N$_2$ & 2.0$\times 10^{-4}$ & 2.0$\times 10^{-4}$ & \\
Mg & 1.0$\times 10^{-9}$ & 1.0$\times 10^{-9}$ & \\
He & 1.0$\times 10^{-1}$ & 1.0$\times 10^{-1}$ & \\
Fe & 1.0$\times 10^{-9}$ & 1.0$\times 10^{-9}$ & \\
Na & 1.0$\times 10^{-9}$ & 1.0$\times 10^{-9}$ & \\
$^{13}$CO & 2.3$\times 10^{-5}$ & 3.8$\times 10^{-5}$ &  \\
H$^{13}$CN & 8.8$\times 10^{-7}$ & 7.7$\times 10^{-9}$ &  \\
C$^{18}$O & 1.6$\times 10^{-6}$ & 1.0$\times 10^{-6}$ &  \\
\hline
  \label{ParentS}
   \end{tabular}
\end{table}

\section{O-type CSEs}

\subsection{Chemical analysis}{\label{M-type-Chem}}

\begin{figure*}
\centering
\begin{minipage}{.5\textwidth}
  \centering
    \includegraphics[width=100mm]{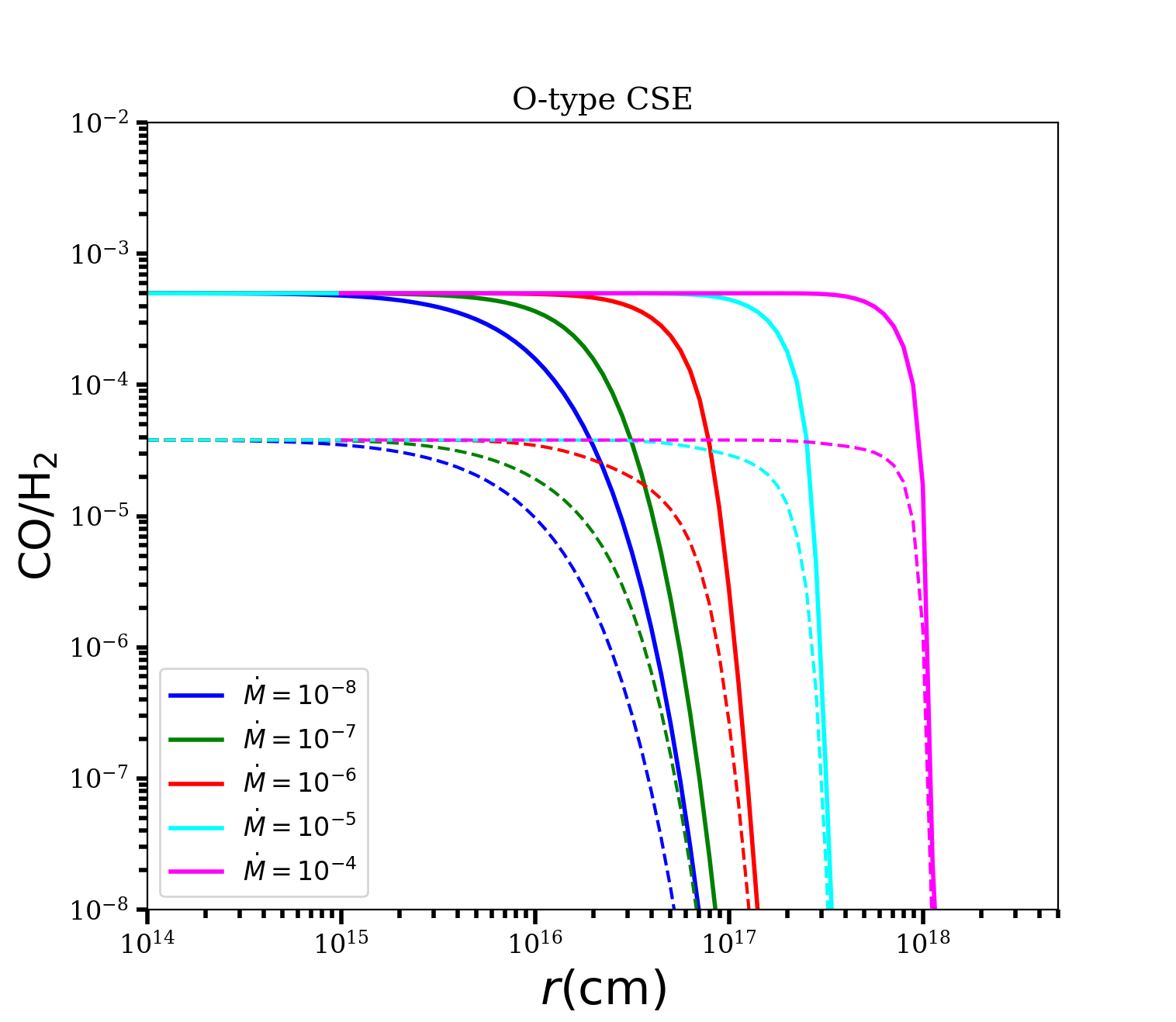}
\end{minipage}%
\begin{minipage}{.5\textwidth}
  \centering
  \includegraphics[width=100mm]{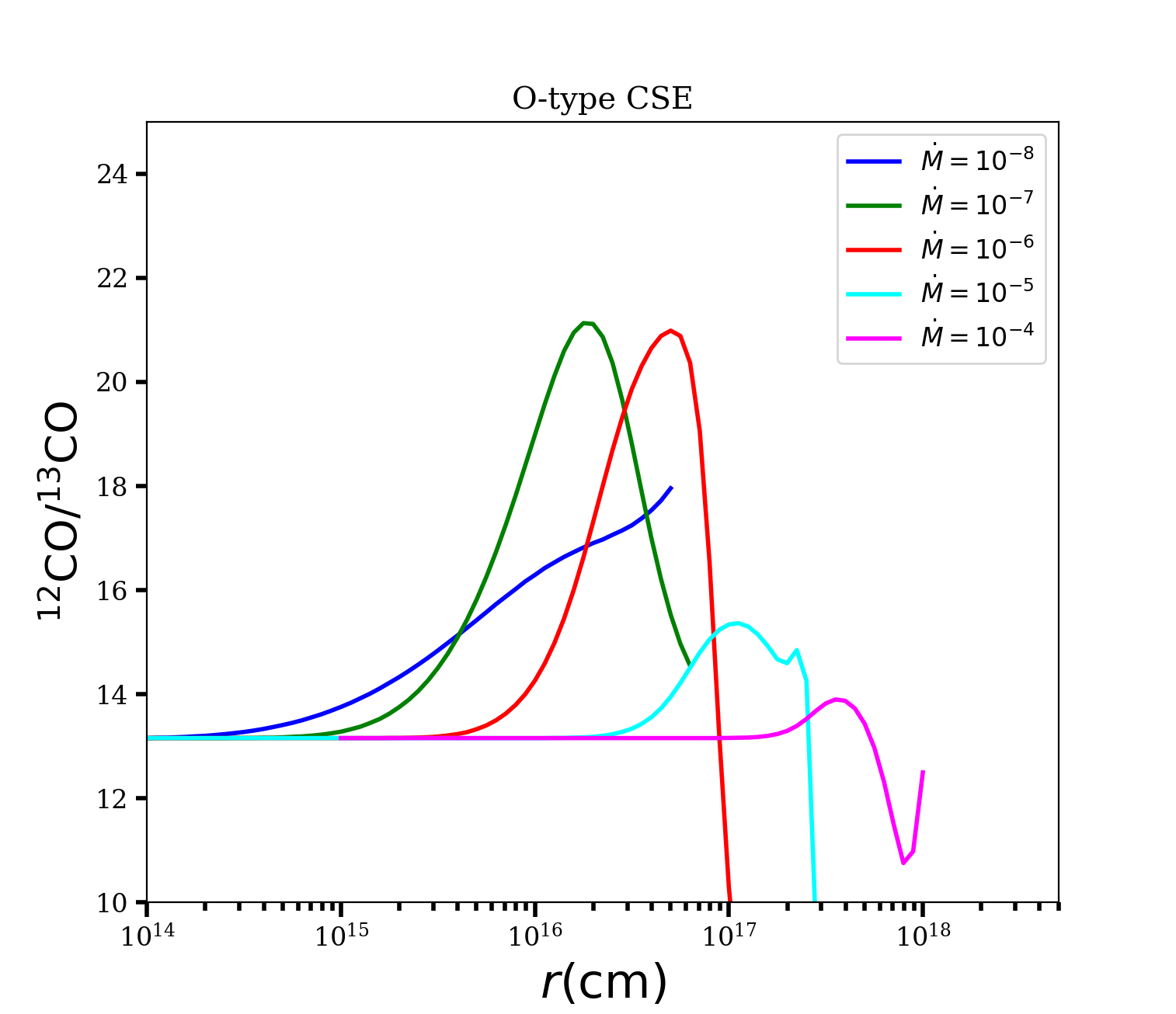}
\end{minipage}
\caption[]{\label{COdist-Mtype} Left:  $\rm ^{12}CO$ (solid lines) and $\rm ^{13}CO$ (dashed lines) abundance profiles for O-type reference models with different mass-loss rates. Right:  $\rm ^{12}CO/^{13}CO$ abundances ratios corresponding to the abundances profiles in the left panel.}
\end{figure*}

\begin{figure*}
\centering
\begin{minipage}{.5\textwidth}
  \centering
  \includegraphics[width=92mm]{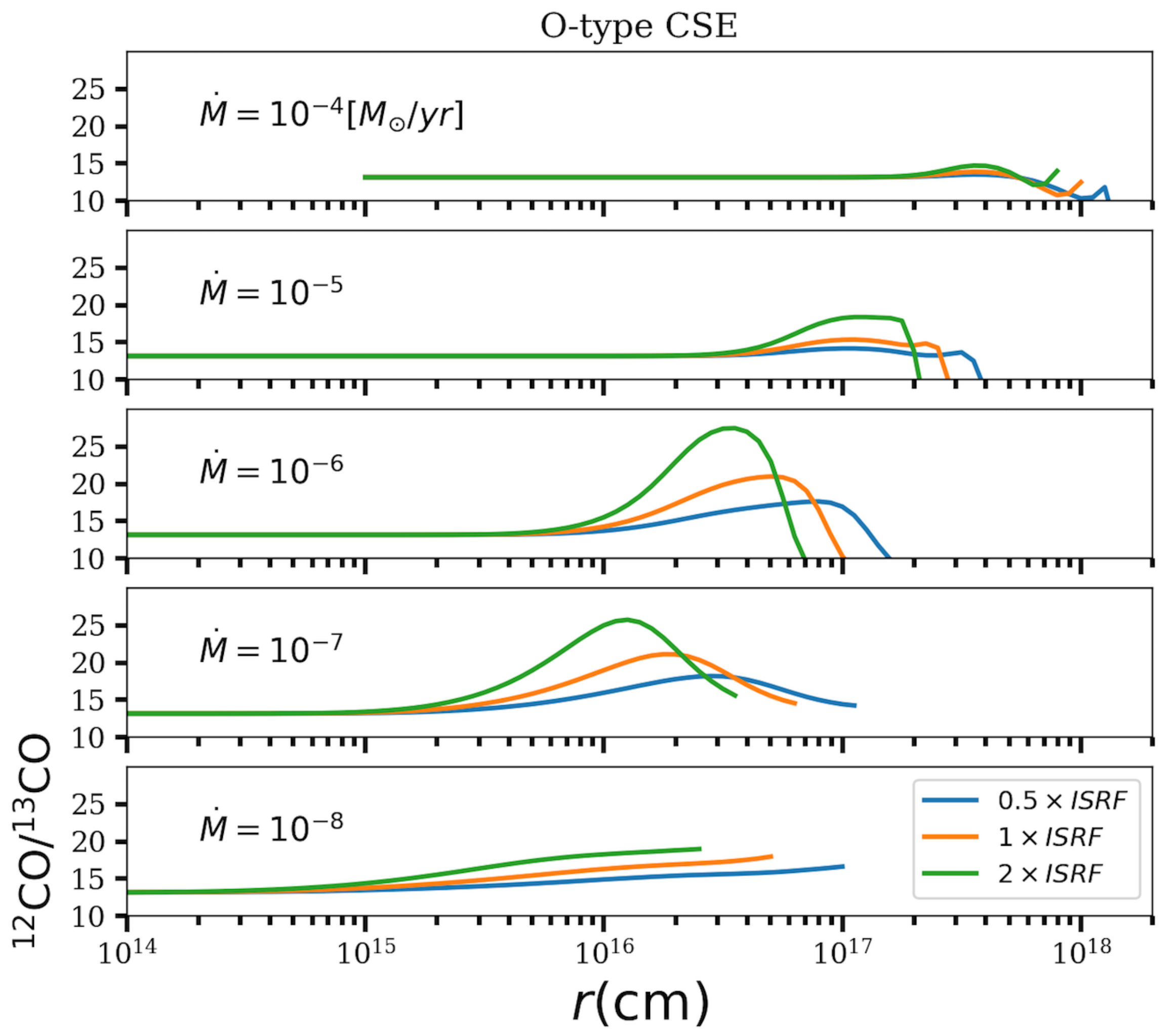}
\end{minipage}%
\begin{minipage}{.5\textwidth}
  \centering
  \includegraphics[width=92mm]{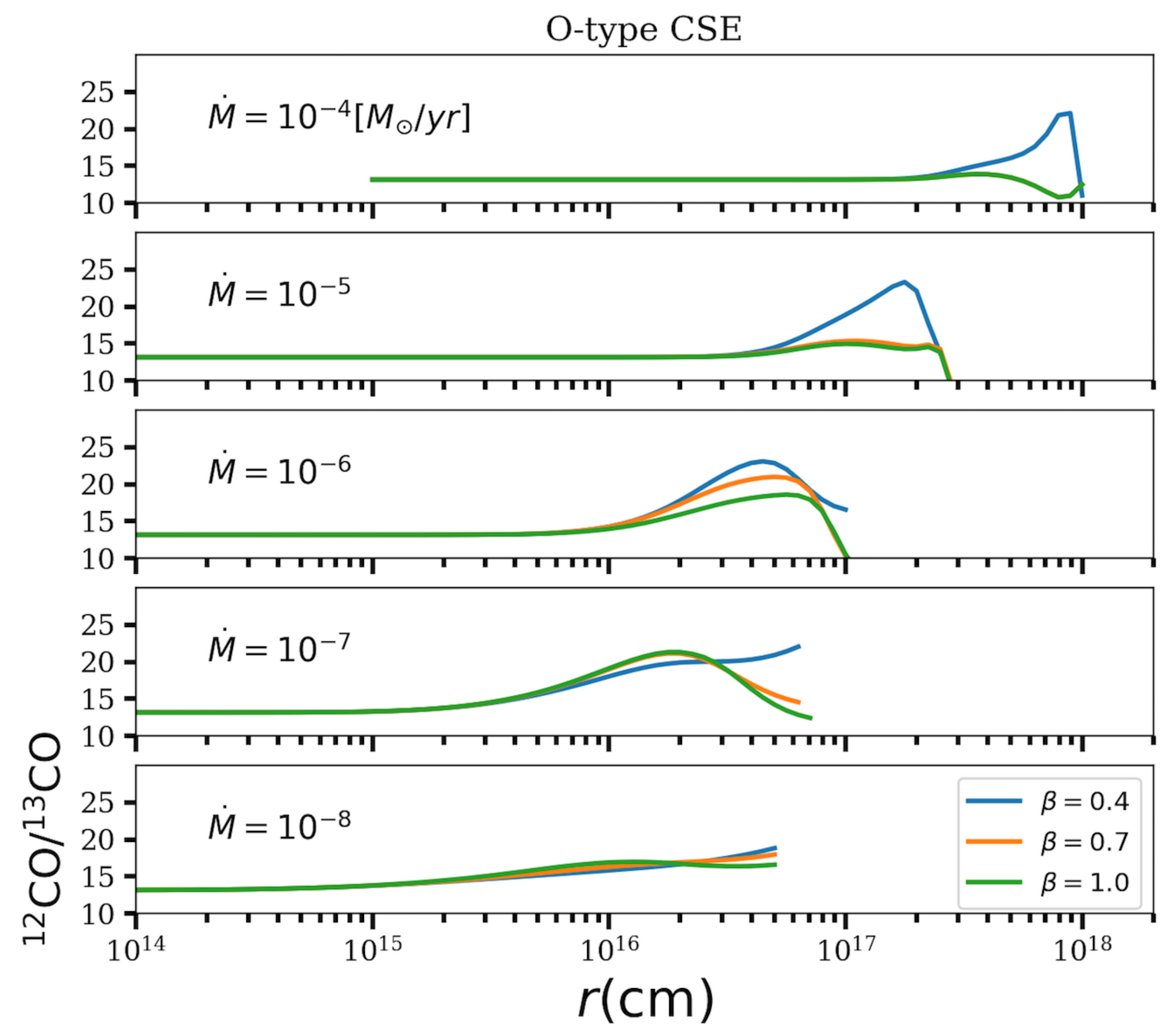}
\end{minipage}
\caption[]{\label{CO-var-M} $\rm ^{12}CO/^{13}CO$ abundance ratios through O-type CSEs with different mass-loss rates under variations of the ISRF intensity (left) and the temperature profile (right).}
\end{figure*}

Here, we presented all the chemical models for the O-type AGB CSEs. The trends are similar to those found for C-type CSEs. Figure~\ref{COdist-Mtype} presents the CO isotopologue abundance distributions and the corresponding ratio for stars with mass-loss rates in the range $10^{-8}<\dot{M}<10^{-4} M_{\odot}$ yr$^{-1}$. 
Table \ref{RCO-O} lists the radii at which two CO isotopologues drop to their initial half-values and the radii where the $^{12}$CO/$^{13}$CO reaches to the maximum deviation from the initial value of 13 for O-type CSEs.
Figure~\ref{CO-var-M} presents the sensitivity of the CO isotopologue abundance ratios to variations of the interstellar radiation field and the gas kinetic temperature for the reference models.

\begin{figure*}
\centering
\begin{minipage}{.5\textwidth}
  \centering
    \includegraphics[width=95mm]{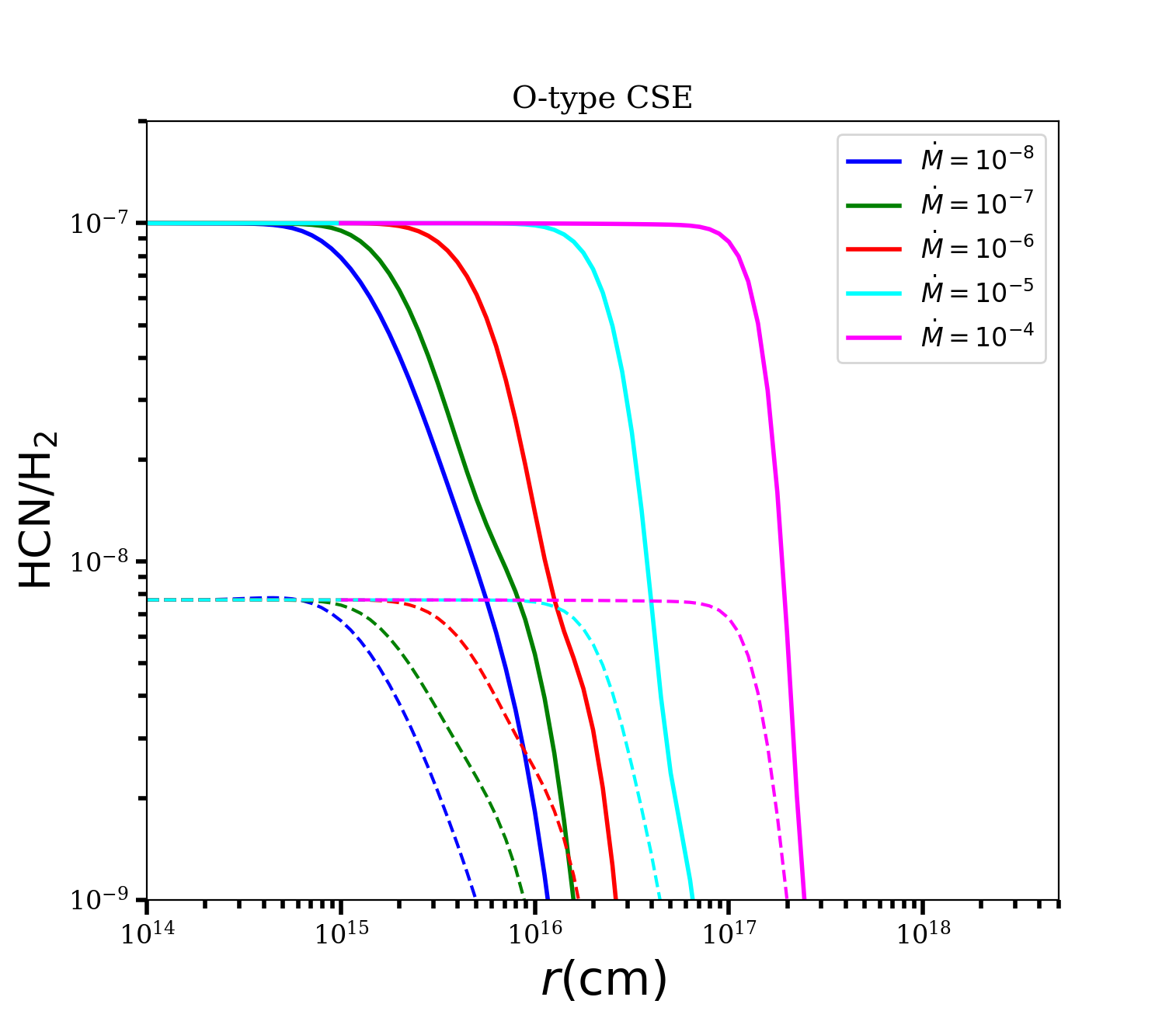}
\end{minipage}%
\begin{minipage}{.5\textwidth}
  \centering
    \includegraphics[width=100mm]{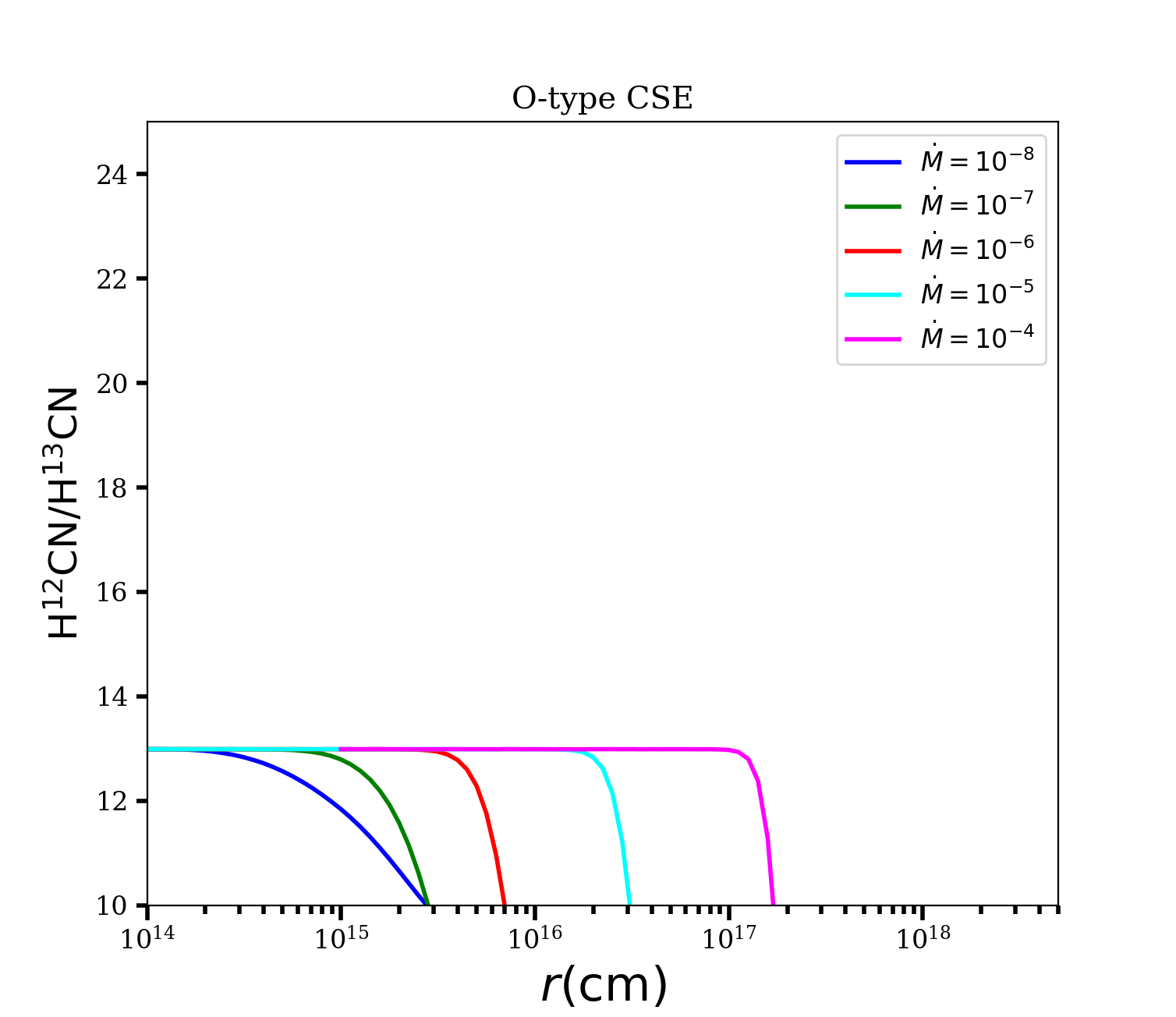}
\end{minipage}
\caption[]{\label{HCN-O} $\rm H^{12}CN$ (solid lines) and $\rm H^{13}CN$ (dashed lines) abundance profiles for O-type CSE models with different mass-loss rates (in units of $M_\odot\, \rm yr^{-1}$) Right:  $\rm H^{12}CN/H^{13}CN$ ratio distributions for the abundance profiles in the left panel.}
\end{figure*}

Figure~\ref{HCN-O} shows the HCN isotopologue abundance distributions and the corresponding ratio for stars with mass-loss rates in the range $10^{-8}<\dot{M}<10^{-4} M_{\odot}$ yr$^{-1}$. As can be seen, the variations of HCN isotopologue ratios are very small in the inner CSEs and become significant only at the outer radii where the H$^{13}$CN abundance drops to very low values and its emission is anyhow undetectable with current observational facilities.


\begin{table}[t]
  \centering
  \setlength{\tabcolsep}{4.0pt}
    \caption{\tiny Results of the chemical model for CO for O-type CSEs.}
  \begin{tabular}{@{} clclclc@{}}
\hline\hline
$\dot{M}$ & $r_{1/2}(^{12}$CO)$^a$& $r_{1/2}(^{13}$CO)$^a$ & $r_{\rm dev_{\rm max}}^b$ & dev$_{\rm max}^c$ \\
 $(M_{\odot}$ yr$^{-1})$  &  (cm) &  (cm) &  (cm) & \\
\hline
$10^{-8}$ &  $6.9 \times 10^{15}$ & $5.6 \times 10^{15}$ & $5.0 \times 10^{16}$ & $36\%$ \\
$10^{-7}$ &  $1.5 \times 10^{16}$ & $1.0 \times 10^{16}$ & $1.8 \times 10^{16}$ & $60\%$\\
$10^{-6}$ & $4.8 \times 10^{16}$ & $3.3 \times 10^{16}$ & $5.0 \times 10^{16}$ & $59\%$ \\
$10^{-5}$ & $1.7 \times 10^{17}$ & $1.6 \times 10^{17}$ & $1.1 \times 10^{17}$ & $17\%$\\
$10^{-4}$ & $7.3 \times 10^{17}$ & $7.5 \times 10^{17}$ & $3.5 \times 10^{17}$ & $6\%$ \\
\hline
 \end{tabular}
\tablefoot{($^a$) Radius at which the CO isotopologue abundances have reached half of their initial values; ($^b$) The radius at which the $^{12}$CO/$^{13}$CO abundance ratio has its maximum deviation from its initial value; ($^c$) The maximum deviation of the $^{12}$CO/$^{13}$CO abundance ratio from its initial value.}
 \label{RCO-O}
\end{table}


\subsection{Radiative transfer analysis}{\label{M-type-RT}}

\begin{figure*}
\centering
\begin{minipage}{.5\textwidth}
  \centering
  \includegraphics[width=100mm]{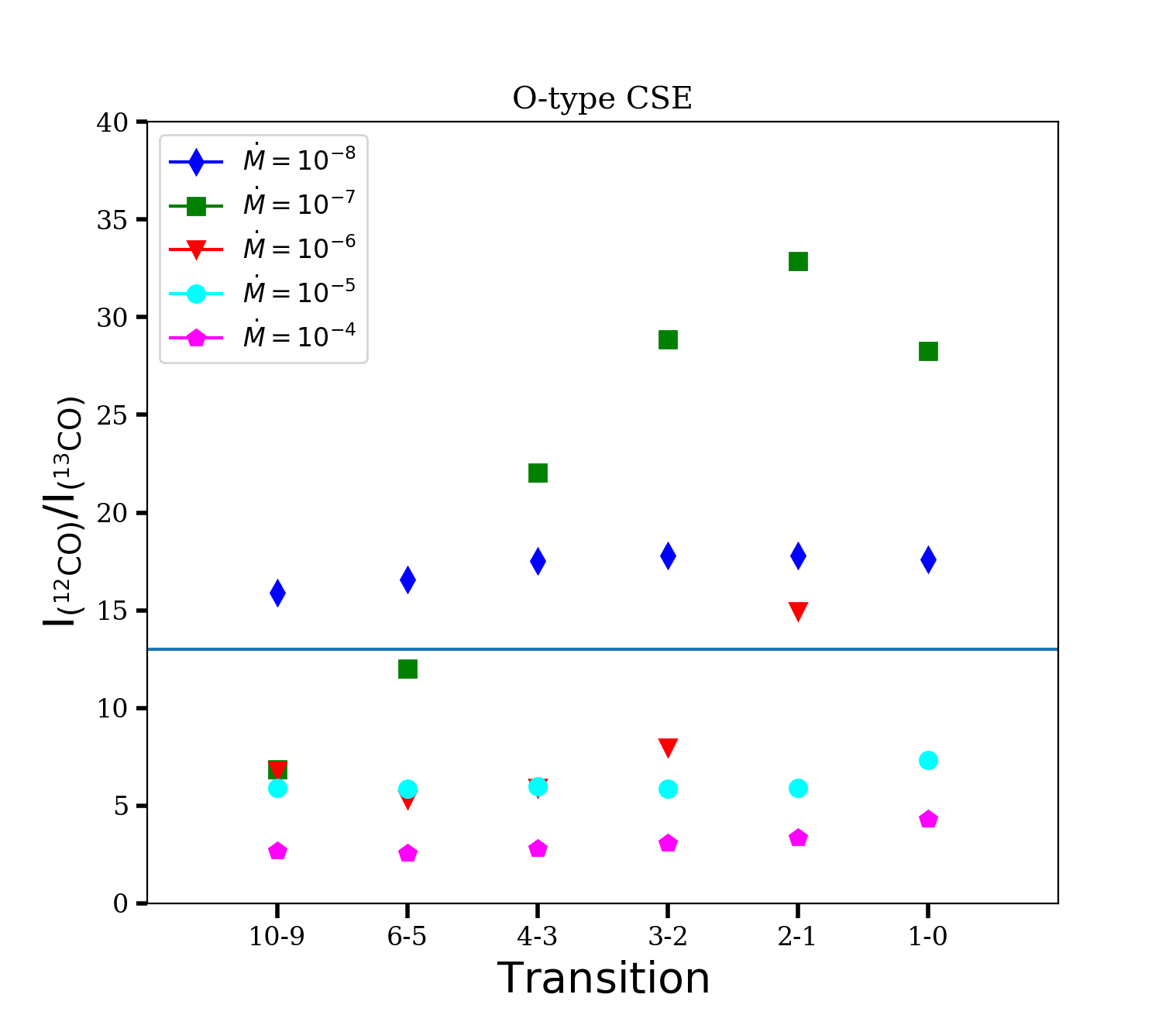}
\end{minipage}%
\begin{minipage}{.5\textwidth}
  \centering
  \includegraphics[width=100mm]{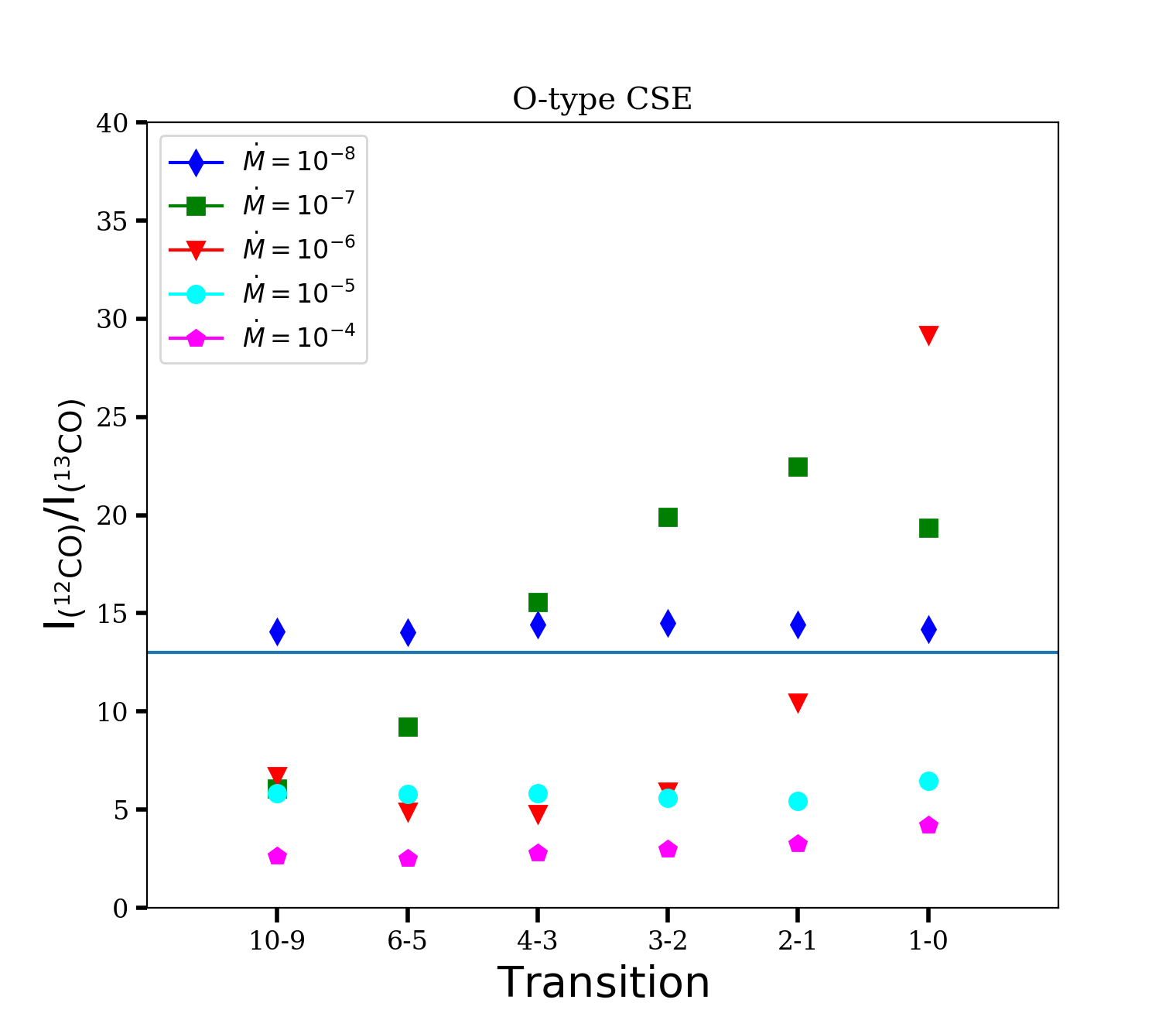}
\end{minipage}
\caption[]{\label{COratio-RT-M} Line intensity ratio $I_{\rm ^{12}CO}/I_{\rm ^{13}CO}$ from RT modeling for reference models with different mass-loss rates with (left) and without (right) considering chemical effects in determining the abundances profiles ((points exceeding values higher than 40 are not plotted). The solid blue lines show the initial adopted abundance ratio of 13 for O-type CSEs.}
\end{figure*}

The line intensity ratio $I_{\rm ^{12}CO}/I_{\rm ^{13}CO}$ for six rotational transitions $J$\,=\,\mbox{1--0}, \mbox{2--1}, \mbox{3--2}, \mbox{4--3}, \mbox{6--5}, and \mbox{10--9} from RT modeling for references models with different mass-loss rates are presented in the left panel of Fig.~\ref{COratio-RT-M}.
Similar to the case for C-type CSEs, there are variations in the line intensity ratios between different rotational transitions for stars with intermediate mass-loss rates ($10^{-7}$ to $10^{-6}$\,$M_{\odot}$\,yr$^{-1}$), while the line intensity ratios for stars with lower and higher mass-loss rates vary little between the transitions. Likewise, the line intensity ratios are shifted to lower values for higher mass-loss rates.

We examined the sensitivity of the line intensity ratios to the abundances distributions from chemical modeling in the same way as for the C-type CSEs. 
Thus, we scaled the $^{12}$CO abundance distribution by a factor of 13 to derive the $^{13}$CO abundance distribution as input in the RT modeling.
The results are presented in the right panel of Fig.~\ref{COratio-RT-M} and variations relative to reference models in percentage are listed under the "No chemistry" label in Table~\ref{variablesM}. 
As for the C-type CSEs, involving chemistry in deriving the molecular abundances mostly affect the line intensity ratio of stars with intermediate mass-loss rates ($10^{-7}$ to $10^{-6}$\,$M_{\odot}$\,yr$^{-1}$).

Table~\ref{variablesM} lists variations of the CO isotopologue line intensity ratios to variations of $T_\star$, $L_\star$, and $R_{\rm in}$ by $\pm20\%$ relative to the results for the reference models. Similar trends as for C-type CSEs are seen. 

\begin{figure*}
  \centering
  \includegraphics[width=90mm]{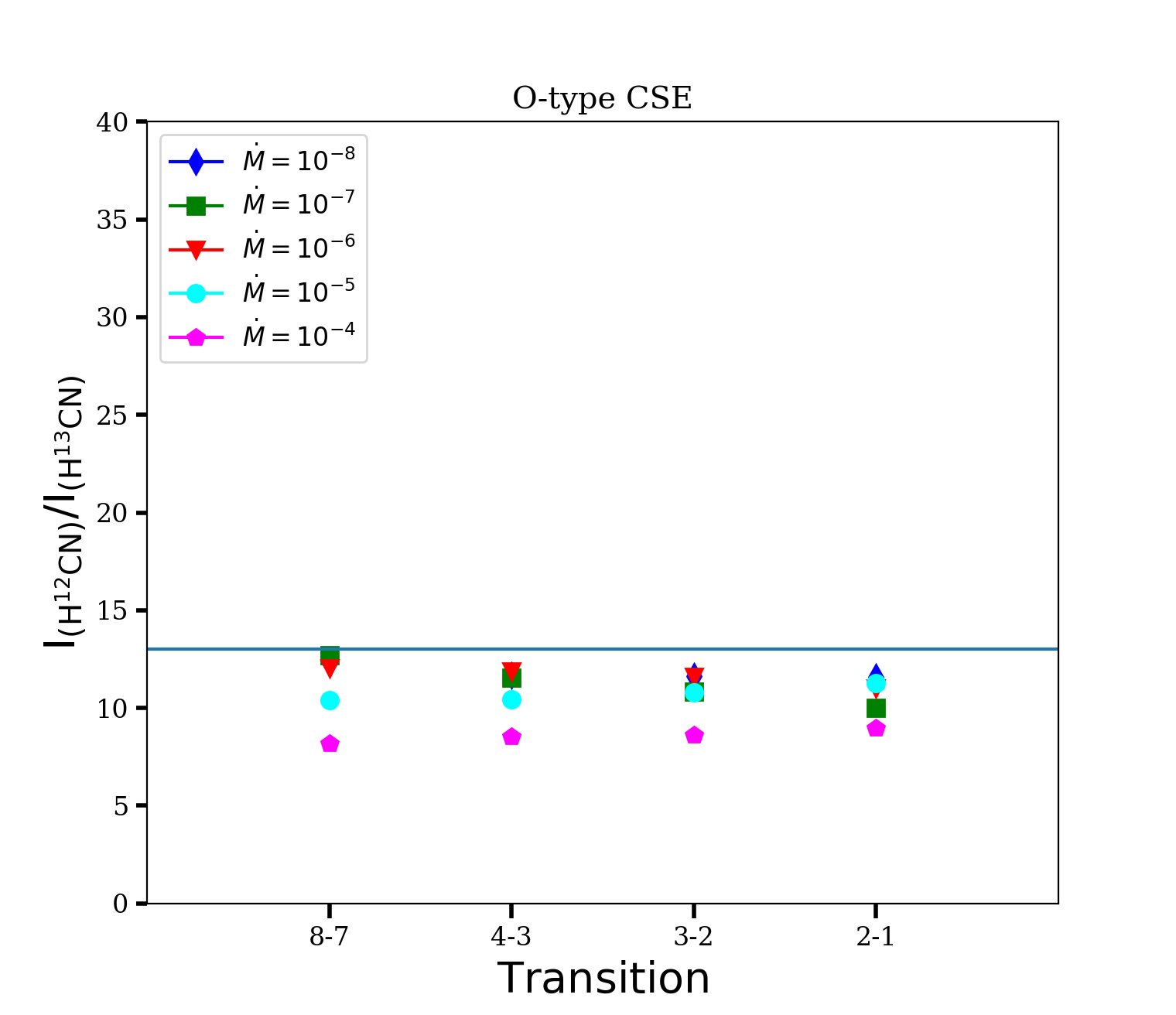}
  \caption[]{
    Line intensity ratio $I_{\rm H^{12}CN}/I_{\rm H^{13}CN}$ from RT modeling for reference models with different mass-loss rates. The solid blue line shows the initial adopted abundance ratio of 13 for O-type CSEs.}
\label{HCN-RT-M}  
\end{figure*}

The line intensity ratio $I_{\rm H^{12}CN}/I_{\rm H^{13}CN}$ for all references models are presented in Fig.~\ref{HCN-RT-M}. The line intensity ratios of different rotational transitions vary little for all models and are shifted to sligthly lower values for higher mass-loss rate stars due to optical depth effects. The effect is much smaller here than in the case of the C-type CSEs.

\begin{table*}[t]
  \centering
  \setlength{\tabcolsep}{5pt}
  \renewcommand{\arraystretch}{1.5}
    \caption{\tiny Change in the $^{12}$CO/$^{13}$CO line intensity ratio compared to the reference models when varying different input parameters of the RT models for O-type CSEs.}
  \begin{tabular}{@{}|c|c|cc|cc|cc|cc|cc|cc|@{}}
\hline
$\dot{M}$ & Variable &  \multicolumn{12}{c}{Variation of $I_{^{12}\rm CO}/I_{^{13}\rm CO}$ relative to reference models} \vline\\
\hline
    &   & \multicolumn{2}{c} { ${J=1\rightarrow0}$} & \multicolumn{2}{c}{${2\rightarrow1}$}  &  \multicolumn{2}{c}{${3\rightarrow2}$} &  \multicolumn{2}{c}{${4\rightarrow3}$}  & \multicolumn{2}{c} {${6\rightarrow5}$}  & \multicolumn{2}{c} {${10\rightarrow9}$}  \vline\\
 
  & & \multicolumn{2}{c} {$E_{\rm up}$=5 (K)} & \multicolumn{2}{c}{$E_{\rm up}$=17}  & \multicolumn{2}{c}{$E_{\rm up}$=33} & \multicolumn{2}{c}{$E_{\rm up}$=55} & \multicolumn{2}{c}{$E_{\rm up}$=116} & \multicolumn{2}{c}{$E_{\rm up}$=304} \vline \\   
\hline
& $T_\star$ & $-20\%$ &$ +20\%$ & $-20\%$ &$ +20\%$& $-20\%$ &$ +20\%$& $-20\%$ &$ +20\%$& $-20\%$ &$ +20\%$ & $-20\%$ &$ +20\%$\\
\hline
$10^{- 8 }$&& 0 \%& 1 \%& 0 \%& 0 \%& 1 \%& 0 \%& 2 \%& -1 \%& 3 \%& -2 \%& 2 \%& -3 \%\\
$10^{- 7 }$&& -5 \%& 5 \%& 1 \%& -3 \%& 10 \%& -10 \%& 16 \%& -13 \%& 17 \%& -11 \%& 0 \%& 2 \%\\
$10^{- 6 }$&& 16 \%& -11 \%& 7 \%& -4 \%& 1 \%& 0 \%& -4 \%& 4 \%& -7 \%& 6 \%& -7 \%& 5 \%\\
$10^{- 5 }$&& 0 \%& -1 \%& 2 \%& 1 \%& 0 \%& 1 \%& 0 \%& 0 \%& -1 \%& 0 \%& 0 \%& 0 \%\\
$10^{- 4 }$&& 2 \%& 1 \%& 0 \%& -1 \%& 0 \%& -2 \%& 1 \%& 0 \%& 1 \%& 1 \%& 0 \%& 0 \%\\
\hline
& $L_\star$ & $-20\%$ &$ +20\%$ & $-20\%$ &$ +20\%$& $-20\%$ &$ +20\%$& $-20\%$ &$ +20\%$& $-20\%$ &$ +20\%$ & $-20\%$ &$ +20\%$\\
\hline
$10^{- 8 }$&& 0 \%& 0 \%& 0 \%& 0 \%& 0 \%& 0 \%& -1 \%& 1 \%& -1 \%& 1 \%& -2 \%& 1 \%\\
$10^{- 7 }$&& 3 \%& -2 \%& -1 \%& 0 \%& -5 \%& 4 \%& -7 \%& 6 \%& -7 \%& 6 \%& 0 \%& 0 \%\\
$10^{- 6 }$&& -7 \%& 5 \%& -2 \%& 2 \%& 1 \%& 0 \%& 2 \%& -1 \%& 3 \%& -2 \%& 3 \%& -2 \%\\
$10^{- 5 }$&& 0 \%& -1 \%& 2 \%& 1 \%& 0 \%& 0 \%& -1 \%& 0 \%& -1 \%& 0 \%& -1 \%& 0 \%\\
$10^{- 4 }$&& -11 \%& 3 \%& -6 \%& 3 \%& -2 \%& -1 \%& 1 \%& 1 \%& 1 \%& 0 \%& 0 \%& 0 \%\\
\hline
 &  $R_{in}$ & $-20\%$ &$ +20\%$ & $-20\%$ &$ +20\%$& $-20\%$ &$ +20\%$& $-20\%$ &$ +20\%$& $-20\%$ &$ +20\%$ & $-20\%$ &$ +20\%$\\
\hline
$10^{- 8 }$&& 1 \%& 0 \%& 1 \%& 0 \%& 1 \%& 0 \%& 1 \%& 0 \%& 1 \%& 0 \%& -1 \%& 0 \%\\
$10^{- 7 }$&& 4 \%& -2 \%& 4 \%& -3 \%& 2 \%& -2 \%& -2 \%& 1 \%& -7 \%& 5 \%& -8 \%& 6 \%\\
$10^{- 6 }$&& -11 \%& 6 \%& -6 \%& 5 \%& -2 \%& 3 \%& 1 \%& 1 \%& 4 \%& -2 \%& 6 \%& -4 \%\\
$10^{- 5 }$&& 1 \%& 0 \%& 4 \%& -2 \%& 4 \%& -4 \%& 4 \%& -4 \%& 3 \%& -3 \%& 2 \%& -2 \%\\
$10^{- 4 }$&& -1 \%& 5 \%& 5 \%& -1 \%& 1 \%& -1 \%& 1 \%& 0 \%& 1 \%& 0 \%& 0 \%& 0 \%\\
\hline
& No chemistry & &&&&&&&&&&&\\
\hline
$10^{-8}$   & &  $-19\%$ && $-19\%$ && $-18\%$ && $-18\%$ && $-15\%$ && $-11\%$ &\\
$10^{-7}$   & &  $-32\%$ && $-32\%$ && $-31\%$ && $-29\%$ && $-23\%$ && $-11\%$ &\\
$10^{-6}$   & & $-32\%$ && $-30\%$ && $-26\%$ && $-19\%$ && $-8\%$ && $-1\%$ &\\
$10^{-5}$   & & $-12\%$ && $-8\%$ && $-5\%$ && $-2\%$ && $-1\%$ && $-1\%$ &\\
$10^{-4}$   & &  $-3\%$ && $-2\%$ && $-2\%$ && $-1\%$ && $-1\%$ && $-1\%$&\\
\hline
 \end{tabular}
 \label{variablesM}
\end{table*}

\section{Excitation of CO}{\label{TauCO}}

\begin{figure*}
\centering
\begin{minipage}{.5\textwidth}
  \centering
  \includegraphics[width=100mm]{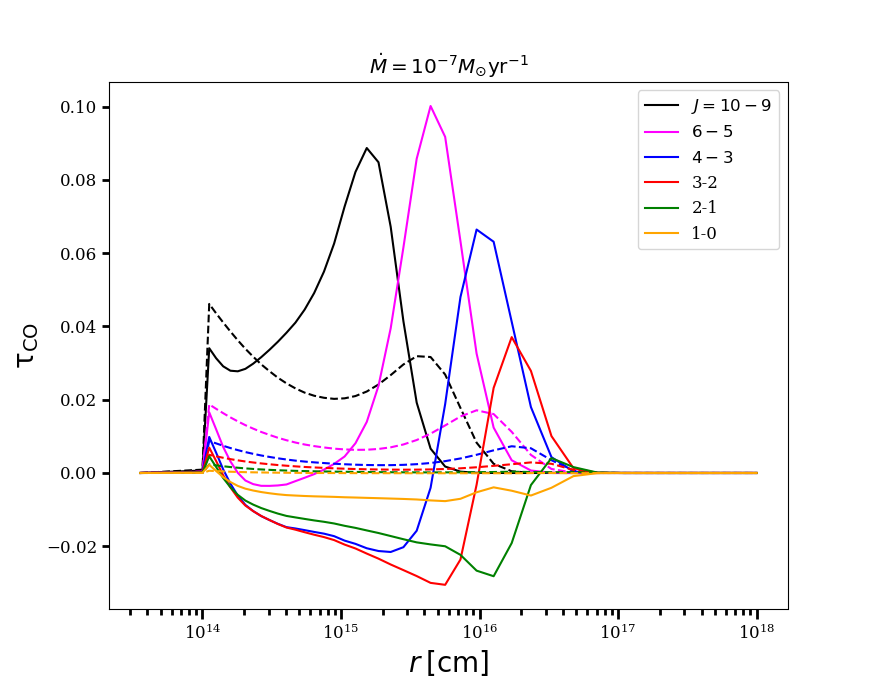}
\end{minipage}%
\begin{minipage}{.5\textwidth}
  \centering
  \includegraphics[width=100mm]{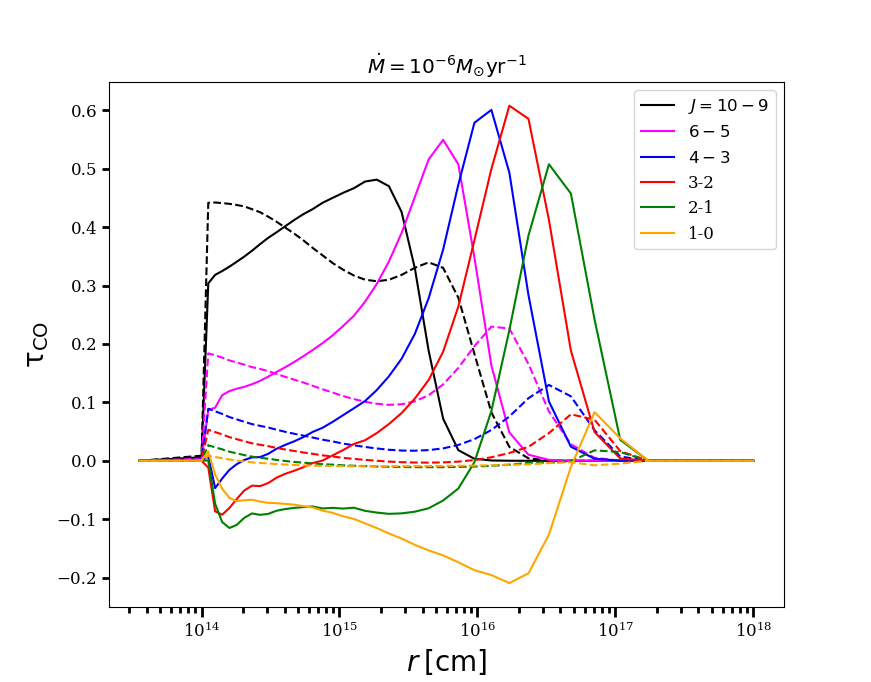}
\end{minipage}
\caption[]{\label{TauCOFig} Tangential line optical depths of $^{12}$CO (solid lines) and scaled $^{13}$CO (by a factor of 50, dashed lines) for C-type CSEs with mass-loss rates $10^{-7}$ (left panel) and $10^{-6}$ (right panel) $M_{\odot}\, \rm yr^{-1}$.}
\end{figure*}

To illustrate differences in excitation between different CO isotopologue rotational transitions, tangential optical depths (optical depth at the systemic velocity along a line of sight intersecting the CSE at various separations from the star) of $^{12}$CO and $^{13}$CO lines for C-type CSEs with mass-loss rates of $10^{-7}$ and $10^{-6} M_{\odot}\, \rm yr^{-1}$ are shown in Fig.\,\ref{TauCOFig}.
As can be seen, the region where the excitation reaches into maximum are not the same for two isotopologues.
The relative importance of radiative and collisional excitation depends on the density (and hence the mass-loss rate). The density at which the excitation goes from being predominantly radiative at lower densities to being predominantly collisional at higher densities is different for the two isotopologues because of their different abundances. This transition density is higher for the lower-abundance species $^{13}$CO.

 \end{appendix}

\end{document}